\documentclass[notitlepage,pra,aps,twocolumn,longbibliography,preprintnumbers,superscriptaddress,nofootnotebib,10pt]{revtex4-2}

\usepackage[dvipsnames]{xcolor}
\definecolor{mediumpersianblue}{rgb}{0.0, 0.4, 0.65}
\definecolor{persianred}{rgb}{0.8, 0.2, 0.2}
\definecolor{darklavender}{rgb}{0.45, 0.31, 0.59}
\usepackage[normalem]{ulem}
\usepackage[colorlinks=true,citecolor=mediumpersianblue,linkcolor=mediumpersianblue, urlcolor=mediumpersianblue]{hyperref}
\renewcommand{\footnoterule}{\vfill\kern -3pt \hrule width 0.4\columnwidth \kern 2.6pt}
\usepackage[utf8]{inputenc}
\usepackage[english]{babel}
\usepackage[T1]{fontenc}
\usepackage{amsmath}
\usepackage{hyperref}[breaklinks=true]
\usepackage{tikz}
\usetikzlibrary{shapes.geometric}
\usetikzlibrary{quantikz}
\usepackage{lipsum}
\usepackage[english]{babel}
\usepackage{graphicx}					
\usepackage[pdftex]{epsfig}
\usepackage{ragged2e}
\usepackage[caption=false]{subfig}
\usepackage{booktabs} 
\usepackage{amsmath,amssymb}
\usepackage{dsfont}
\usepackage{fontawesome}
\usepackage{siunitx}
\usepackage{mathtools}
\usepackage{bm}
\usepackage{multirow}
\usepackage{braket}
\usepackage{amsmath}
\usepackage{amssymb}
\usepackage{amsthm}
\usepackage{soul}
\usepackage{dsfont}
\def\id{\mathds{1}}

\usepackage[dvipsnames]{xcolor}
\usepackage{dsfont}
\def\id{\mathds{1}}
\newcommand{\quotes}[1]{``#1''}

\DeclareMathAlphabet{\mathpzc}{OT1}{pzc}{m}{it}

\newcommand{\X}{\hat X}
\newcommand{\Y}{\hat Y}
\newcommand{\Z}{\hat Z}
\newcommand{\XXZ}{\operatorname{XXZ}}

\def\h{\hat H}
\def\D{\hat D}
\def\v{\hat V}
\def\bfzero{\boldsymbol{0}}
\def\bftheta{\boldsymbol{\theta}}
\def\vqe{\u_{\bftheta}}
\def\vqes{\u_{\bftheta^*}}

\newcommand{\ketbra}[2]{\ket{#1} \!\! \bra{#2}}

\def\u{{\hat U}}

\def\D{\hat D}

\def\RDBI{\hat R}

\def\D{\mathcal D}
\def\Z{\hat Z}

\def\D{\hat D}

\newcommand\del[1]{}
\newtheorem*{theorem*}{Theorem}

\usepackage{mathcomp}
\usepackage{utfsym}

\begin{document}

\title{Double-bracket quantum algorithms for high-fidelity ground state preparation}
\preprint{TIF-UNIMI-2024-6}

\newcommand{\MIaff}{TIF Lab, Dipartimento di Fisica, Universit\`a degli Studi di Milano}
\newcommand{\INFN}{INFN Sezione di Milano, Milan, Italy}
\newcommand{\TII}{Quantum Research Center, Technology Innovation Institute, Abu Dhabi, UAE}
\newcommand{\CERNaff}{European Organization for Nuclear Research (CERN), Geneva 1211, Switzerland}
\newcommand{\ANU}{School of Computing, The Australian National University, Canberra, ACT, Australia}
\newcommand{\UNIGE}{Department of Nuclear and Particle Physics, University of Geneva, Geneva 1211, Switzerland}
\newcommand{\NTU}{School of Physical and Mathematical Sciences, Nanyang Technological
University, 21 Nanyang Link, 637371 Singapore, Republic of Singapore}
\newcommand{\astar}{Institute of High Performance Computing (IHPC), Agency for Science, Technology and Research (A*STAR), 1 Fusionopolis Way, \#16-16 Connexis, Singapore 138632, Singapore}
\newcommand{\EPFL}{Institute of Physics, Ecole Polytechnique Fédérale de
Lausanne (EPFL), Lausanne, Switzerland}
\newcommand{\UFRJ}{Instituto de F\'{i}sica, Universidade Federal do Rio de Janeiro, 
P.O. Box 68528, Rio de Janeiro, Rio de Janeiro 21941-972, Brazil}
\newcommand{\ETHZ}{Institute for Theoretical Studies, ETH Zurich, 8092 Zurich, Switzerland}

\author{Matteo Robbiati}
\thanks{MR, EP, AP, LX, and OK contributed equally}
\affiliation{\CERNaff}
\affiliation{\MIaff}

\author{Edoardo Pedicillo}
\thanks{MR, EP, AP, LX, and OK contributed equally}
\affiliation{\MIaff}
\affiliation{\TII}

\author{Andrea Pasquale}
\thanks{MR, EP, AP, LX, and OK contributed equally}
\affiliation{\MIaff}
\affiliation{\TII}
\affiliation{\INFN}

\author{Li Xiaoyue}
\thanks{MR, EP, AP, LX, and OK contributed equally}
\affiliation{\NTU}

\author{Oriel Kiss}
\thanks{MR, EP, AP, LX, and OK contributed equally}
\affiliation{\CERNaff}
\affiliation{\UNIGE}

\author{Andrew Wright}
\affiliation{\EPFL}

\author{Renato M. S. Farias}
\affiliation{\TII}
\affiliation{\UFRJ}

\author{Khanh Uyen Giang}
\affiliation{\NTU}

\author{Jeongrak Son}
\affiliation{\NTU}

\author{Johannes Kn\"orzer}
\affiliation{\ETHZ}

\author{Siong Thye Goh}
\affiliation{\astar}

\author{Jun Yong Khoo}
\affiliation{\astar}

\author{Nelly H.Y. Ng}
\affiliation{\NTU}

\author{Zo\"{e} Holmes}
\affiliation{\EPFL}

\author{Stefano Carrazza}
\affiliation{\CERNaff}
\affiliation{\MIaff}
\affiliation{\TII}
\affiliation{\INFN}

\author{Marek Gluza}
\email{marekludwik.gluza@ntu.edu.sg}
\affiliation{\NTU}

\begin{abstract}
Ground state preparation is a central application for quantum computers but remains challenging in practice. In this work, we quantitatively investigate the performance and gate counts of double-bracket quantum algorithms (DBQAs) for ground state preparation. We propose a practical strategy in which DBQAs refine initial state preparation circuits, and we compile them for Heisenberg chains using controlled-Z and single-qubit gates. Warm-started DBQAs consistently improve both the energy and ground-state fidelity relative to the initial states provided by variational ans\"atze, indicating that DBQAs offer an effective unitary synthesis method. To demonstrate compatibility with near-term hardware, we executed a proof-of-concept example on IBM devices. With error mitigation, we observed a statistically significant improvement over the corresponding warm-start circuit. Furthermore, numerical emulations for the same system size indicate that executing DBQAs on Quantinuum's hardware could achieve similar cost-function gains without requiring error mitigation. These findings suggest that DBQAs are a promising approach for enhancing ground-state approximations on near-term quantum devices.
\end{abstract}

\maketitle

\section{Introduction}\label{sec:introduction} Approximating the ground state of a target Hamiltonian is a challenging problem that has been widely investigated in quantum and classical computing.
Quantum phase estimation~\cite{Dong_2022, PRXQuantum.3.010318, ge2019faster,EFTQC_practice, grimsley2019adaptive} and quantum signal processing~\cite{PRXQuantum.6.020327,Lin2020nearoptimalground} are expected to be the go-to methods for computing ground state energies in the fault-tolerant era, but the circuit depths they require are prohibitive on near-term devices.
Instead, in the near term, parametrized circuits are widely used for cost function optimization~\cite{KishorRevModPhys.94.015004, cerezo2021variationalReview, tilly2022variationalReview}, but lack convergence guarantees and often encounter numerous optimization barriers~\cite{anschuetz2022beyond, McClean_2018, klieschVQE, stilck2021limitations, Larocca2024, cerezo2023does}.

Recently, double-bracket quantum algorithms (DBQAs) have been introduced for Hamiltonian diagonalization~\cite{Gluza_2024}, imaginary-time evolution~\cite{gluza2024double,zander2025role}, and quantum signal processing~\cite{suzuki2025double}.
These methods are grounded in Riemannian gradient descent~\cite{helmke_moore_optimization}, which provides a rigorous basis for their performance and convergence guarantees~\cite{gluza2024double}.
The DBQA framework has also enabled a derivation of Grover’s algorithm directly from optimization principles on Riemannian manifolds~\cite{suzuki2025grover}.
Taken together, these results indicate that DBQAs provide a unifying framework for ground state preparation. However, to be practically viable, the gate cost must be small.
In this manuscript, we investigate the potential of DBQAs for preparing ground states on early fault-tolerant quantum devices \cite{early_FTQC}. 
Unlike quantum phase estimation~\cite{ge2019faster} or block-encoding formulations of quantum signal processing~\cite{Lin2020nearoptimalground}, DBQAs can be implemented as near-term quantum circuits without the need for auxiliary qubits.
Crucially, unlike parametrized circuits, DBQAs involve classical optimization only for performance improvement and can be analytically guaranteed to converge~\cite{helmke_moore_optimization,moore1994numerical,smith1993geometric,gluza2024double}.
\begin{figure}[t!]
  \centering
    \includegraphics[width=1\linewidth]{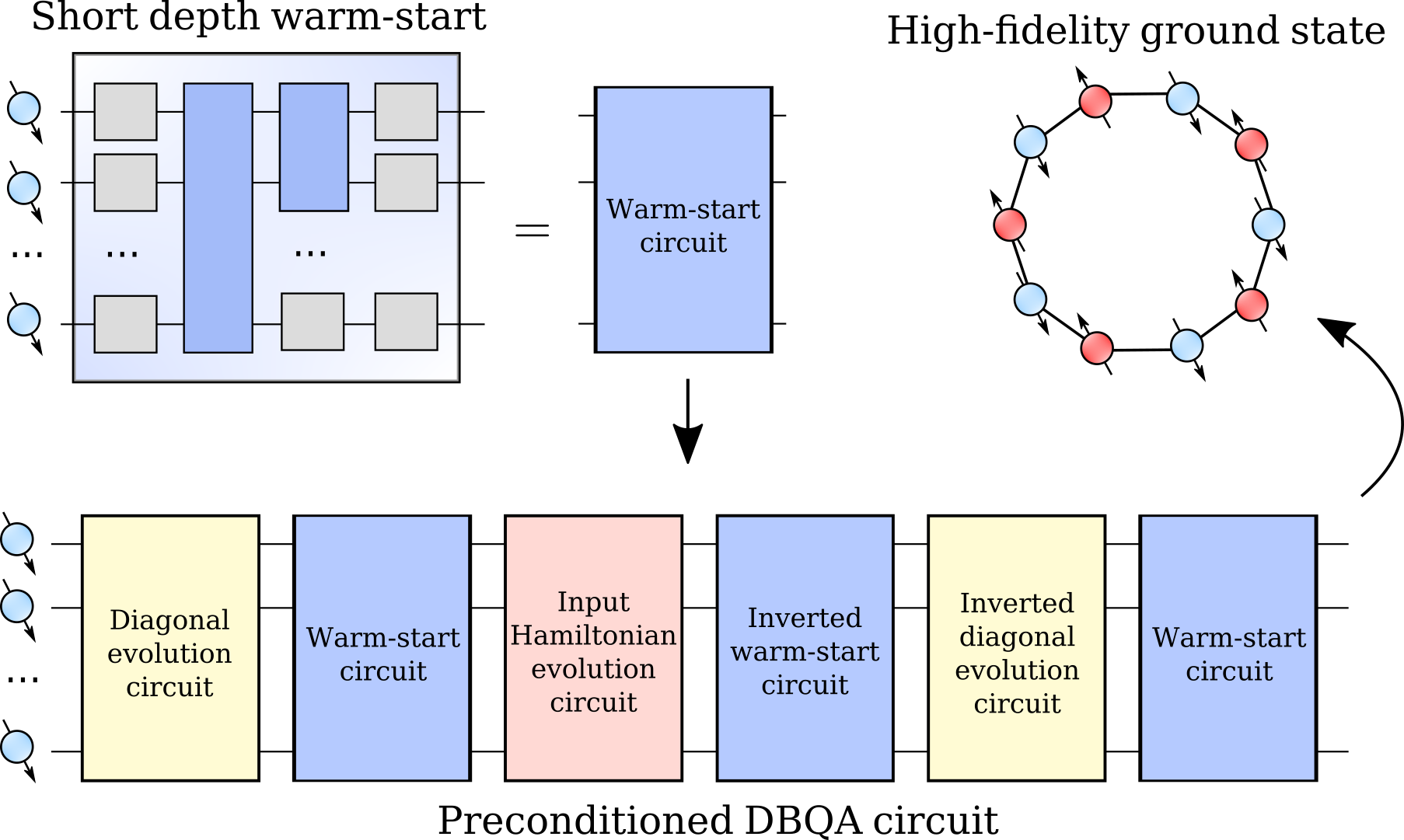} %
	\caption{
 We propose a two-stage ground state preparation protocol: first, apply a relatively short-depth warm-start circuit; second, apply a DBQA circuit to further the ground state preparation fidelity. }
    \label{fig:scheme}
\end{figure}

In this work, as schematically shown in Fig.~\ref{fig:scheme}, we propose a two-stage ground state preparation protocol: first, apply an existing state preparation method that uses relatively short-depth circuits; second, apply a DBQA to further improve the state preparation. 
We selected the variational quantum eigensolver (VQE) as our initialization method not only for its low overhead and ease of unitary retrieval, but more importantly to emphasize that DBQA can refine even limited approximations, such as those achievable with short-depth circuits on near-term hardware.
However, we stress that our method is largely agnostic to the choice of initialization method and there exist many potential candidates depending on the problem, including: parametrized circuits~\cite{Peruzzo_2014}, the Hamiltonian variational ansatz~\cite{bosse_Heisenberg_2022, kattemolle_vanwezel_2022,wecker2015progress}, Hartree-Fock circuits~\cite{arute2020hartree}, tensor networks~\cite{DMRG_White,DMRG_Wilson}, quantum imaginary-time evolution~\cite{Motta2020,nishi2021, mcArdle2019,avqite}, coupled cluster~\cite{Fomichev_initial_state}, dissipative \cite{dissipation_cirac} and qubitization~\cite{Dong_2022, PRXQuantum.3.010318, ge2019faster} approaches. 

To demonstrate the effectiveness of our method, we perform numerical simulations of DBQA for Heisenberg chains. 
We investigate the dependence of DBQA circuit depths on the quality of the initial approximation of the ground state.
As a specific example, for $L=10$ qubits, a single DBQA step (with circuit depth of $50$-$100$ CZ gates per qubit) can improve the energy estimation by an order of magnitude when initialized through a parametrized circuit with energy closer to the ground state than the first excited state.
Finally, we validate our approach with a 10-qubit experiment on IBM's quantum hardware and perform numerical emulation of Quantinuum's hardware.

\vspace{1.5mm}
\section{Double-bracket diagonal-operator iterations (DB-DOI)}
DBQAs are based on differential equations called double-bracket flows (DBFs), which have applications in quantum many-body systems~\cite{flow_equation,PhysRevD.48.5863,wegner2006flow,Kehrein:2006ti} and control theory ~\cite{helmke_moore_optimization}. 
We discuss two DBFs that play a role in quantum computing.
We will summarize the DBQA framework as the conjunction of two approximations that produce quantum circuits that approximate the ground state of an input Hamiltonian $\h_0$.
The Brockett's DBF~\cite{BROCKETT199179} is given by
\begin{align}
    \partial_\ell\h_0(\ell) = [ [\hat D,\h_0(\ell)],\h_0(\ell)]
    \label{eq:DBF H}
\end{align}
and we set $\h_0(0)=\h_0$.
This equation preserves the spectrum of $\h_0$ if the operator $\hat D$ is Hermitian.
If $\D$ is diagonal and non-degenerate, then $\h_0(\infty)$ will be diagonal too.
Moreover, the spectrum of $\h_0(\infty)$ will be sorted according to the ordering of the eigenvalues of $\D$, which is an important feature when using Brockett's DBF for ground state preparation~\cite{helmke_moore_optimization}.

The first approximation to the DBF made in the DBQA framework is a first-order discretization of the differential equation Eq.~\eqref{eq:DBF H}.
We set $s$ to be a small Euler-step duration and arrive at a discrete sequence of Hamiltonians defined recursively by
\begin{align}
    \h_{k+1} = e^{s[\D, \h_k]}\h_ke^{-s[\D, \h_k]}\ .
    \label{def DBI}
\end{align}
We notice that
\begin{align}
\h_{k+1} = \h_k +s[[\D, \h_k],\h_k] +O(s^2)\ ,
\end{align}
which links Eqs.~\eqref{eq:DBF H} and \eqref{def DBI}.
Ref.~\cite{Gluza_2024} explored the convergence of this sequence to the DBF in the limit $s\rightarrow 0$ and proposed to additionally vary the operator $\D$ in every step $k$ to reduce the number of steps of the algorithm.
Ref.~\cite{helmke_moore_optimization} details that the sequence $\h_k$ converges to a diagonal fixed point, similarly as the DBF.
This means that 
\begin{align}
    \hat R_k =e^{-s[\D, \h_0]}e^{-s[\D, \h_1]}\ldots e^{-s[\D, \h_k]}
    \label{eq:RkDBI}
\end{align}
converges to a diagonalization unitary.
This realizes the first benefit of the DBQA framework: We obtain an analytical pathway to perform unitary synthesis of circuits approximating eigenstates.
Let us remark that Brockett's DBF is a Riemannian gradient flow, which indicates its local optimality and predicts an exponentially fast convergence when initialized in the basin of attraction of the fixed point~\cite{helmke_moore_optimization,moore1994numerical,smith1993geometric}.

The second approximation step in the DBQA framework is to approximate exponentials of commutators by a product of exponentials.
Specifically we can use that~\cite{Gluza_2024}\begin{align}
    e^{-s[\D, \h_k]} =e^{-i\sqrt{s}\h_k} e^{i\sqrt{s}\D}e^{i\sqrt{s}\h_k}e^{-i\sqrt{s}\D} + O(s^{3/2})\ .
    \label{basic gc eq}
\end{align}
The benefit of this approximation is that each of the 4 exponentials in the product can be performed on the quantum computer.
For example, $\D$ could be a magnetic field and we assume that $\h_0$ is a local Hamiltonian.
We complete the use of this approximation by redefining the iteration of Hamiltonians $\h_k$ to be defined by
\begin{align}
    \h_{k+1} = e^{i\sqrt{s}\D}e^{-i\sqrt{s}\h_k}e^{-i\sqrt{s}\D} \h_k e^{i\sqrt{s}\D}e^{i\sqrt{s}\h_k}e^{-i\sqrt{s}\D}
    \label{eq dropping eith}
\end{align}
and note that we have used $e^{i\sqrt{s}\h_k} \h_k e^{-i\sqrt{s}\h_k} = \h_k$.

To summarize the above, we track the unitary operations that will be applied to the quantum state to obtain an eigenstate approximation. For this, let us denote by $\hat U_k$ the unitary that rotates the initial Hamiltonian to the Hamiltonian after the $k$-th iteration step by 
\begin{align}
    \h_k = \hat U_k^\dagger \h_0 \hat U_k\ .
    \label{h DOI}
\end{align}
We then can use that $e^{it\h_k} = U_k^\dagger e^{it\h_0} U_k$.
With this, we arrive at the following recursive sequence of unitaries $\hat U_k$ which starts with $\hat U_0 = \id$ and proceeds by
\begin{align}
    \hat U_{k+1} = \hat U_ke^{i\sqrt{s}\D}{\hat U_{k}}^\dagger e^{i\sqrt{s}\h_0} \hat U_k e^{-i\sqrt{s}\D} \ .
    \label{db-doi}
\end{align}
In the limit $s\rightarrow0$, we will obtain convergence to Brockett's DBF~\cite{Gluza_2024}, with the difference to Eq.~\eqref{eq:DBF H} that the recursive definition involves only evolutions under $\D$ and $\h_0$, and we obtain an actionable quantum algorithm.
We will call the unitaries in Eq.~\eqref{db-doi} \textit{double-bracket diagonal-operator iterations} (DB-DOI).

Eq.~\eqref{eq:DBF H} pertains to the Hamiltonian and, based on rotating it, it arrives at a unitary that can be applied to computational basis states to obtain approximations of eigenstates.
We next consider a related DBF procedure that is instead focused on the input state rather than the Hamiltonian. Given an initial state $\ket{\psi_0}$, we can consider the following DBF
\begin{align}
    \partial_\tau\Psi(\tau) = [ [\Psi(\tau),\h_0],\Psi(\tau)]\ ,
    \label{eq:DBF ITE}
\end{align}
where $\Psi(0) = \ket{\psi_0}\bra{\psi_0}$.
This is again a Brockett's DBF as in Eq.~\eqref{eq:DBF H}, but in the Schr\"odinger picture because the density matrix of the input state is being rotated and the Hamiltonian plays the role of the constant target matrix $\D$. If we define imaginary-time evolution (ITE) by
\begin{align}
    \ket{\psi(\tau)} =\frac{ e^{-\tau\h_0}\ket{\psi_0}}{\|e^{-\tau\h_0}\ket{\psi_0}\|}
\end{align}
then we have precisely $\Psi(\tau)=\ketbra{\psi(\tau)}{\psi(\tau)} $~\cite{gluza2024double}.
This observation has led to the proposal of a new DBQA called double-bracket quantum imaginary-time evolution (DB-QITE)~\cite{gluza2024double}, which we will next summarize in comparison to the above DB-DOI discussion.

The first approximation step of the DBQA framework leads us to
\begin{align}
    \ket{\psi(\tau)} =e^{s[ \ket{\psi_0}\bra{\psi_0},\h_0]}\ket{\psi_0} + O(\tau^2)\ .
\end{align}
This has the important benefit of providing an analytical ansatz for unitary synthesis that implements ITE, which a priori involves a non-unitary propagation.
Next, after performing the second approximation step of the DBQA framework, \emph{i.e.} using a product formula approximation to the exponential of a commutator, we arrive at the recursive relation
\begin{align}
    \hat U_{k+1} = e^{i\sqrt{s}\h_0}{\hat U_{k}}^\dagger e^{i\sqrt{s}\ket{\psi_0}\bra{\psi_0} }\hat U_k e^{-i\sqrt{s}\h_0} \hat U_k\ 
    \label{db-dbqite}
\end{align}
which is rooted in $\hat U_0 = \id$.
Ref.~\cite{gluza2024double} shows that Eq.~\eqref{db-dbqite} allows to reduce the energy achieved by any circuit $\hat U_0\neq \id$ by repeating it three times, which without the DBQA framework appears to be difficult to derive~\cite{AnshuImproved}.

\subsection{Contrasting DB-DOI with DB-QITE}
Both DB-DOI and DB-QITE qualify as double-bracket quantum algorithms (DBQAs), while their distinction lies in formulation: DB-DOI is naturally expressed in the Heisenberg picture, whereas DB-QITE is defined in the Schr\"odinger picture.
The two can be related by observing that $\D$ and $\h_0$ in DB-DOI play a role analogous to $\h_0$ and $\ketbra{0}{0}$ in DB-QITE. In particular, $\D$ in DB-DOI and $\h_0$ in DB-QITE serve as Brockett’s target matrices, which define the linear cost function in the Riemannian gradient descent formulation of the double-bracket flow (DBF). In contrast, $\h_0$ in DB-DOI and $\ketbra{0}{0}$ in DB-QITE are the dynamical variables that are iteratively transformed towards the desired fixed point.

A priori, the fixed point of DB-DOI does not directly address ground-state preparation. However, Brockett’s DBFs exhibit a sorting property: generically, the spectrum of the limiting operator $H_\infty$ is ordered according to the eigenvalue ordering encoded in $\D$. Consequently, the ground state of $\D$ corresponds to the ground state of $\h_0$ upon convergence of DB-DOI. In contrast, the fixed point of DB-QITE is explicitly the ground state itself. Both algorithms originate from Brockett’s DBF, a Riemannian steepest-descent flow~\cite{schulte2010gradient}. For discussions in the context of quantum computing, see Refs.~\cite{suzuki2025grover,gluza2024double,suzuki2025double,zander2025role}, and for a general treatment of DBFs, see the monograph~\cite{helmke_moore_optimization}. The quantum algorithmic prescription for both DB-DOI [Eq.~\eqref{db-doi}] and DB-QITE [Eq.~\eqref{db-dbqite}] are a recursive unitary synthesis.

 The input Hamiltonian $\h_0$ is the variable in DB-DOI and is typically full rank. 
 In contrast, for DB-QITE the variable $\ketbra{0}{0}$ is rank $1$.
Thus, this makes Brockett's DBF sorting property far more striking: instead of sorting the full spectrum, we only shift the energy of the pure state to match the ground state.
On the other hand, it can be beneficial in other cases that DB-DOI can also converge to excited eigenstates by initializing in an excited state of $\D$.
To achieve that, DB-QITE would need to perform typical scanning of the folding spectrum \emph{i.e.} consider $\h_0'(\lambda) = (\h_0-\lambda)^2$ with various $\lambda$ because then the ground state of $\h_0'(\lambda)$ will be the eigenstate of $\h_0$ with eigenvalue closest to $\lambda$.

It is no coincidence that DB-QITE involves $e^{-i\sqrt{s}\ketbra{0}{0}}$, where a similar term appears in Grover's algorithm: the latter can be derived by the two steps of the DBQA framework~\cite{suzuki2025grover}.
This type of operation has recently led to a new unitary synthesis for quantum signal processing which does not involve post-selection.
For both these insights, the rank-1 property of the variable has been key.
It might be that when the DBQA involves $\D$, which is full rank, it is harder to uncover new properties.

It is crucial to note that it is possible to express $e^{-i\sqrt{s}\ketbra{0}{0}}$ in terms of $\mathrm{CZ}$ and single-qubit gates with a linear scaling in the number of qubits.
This holds even for architectures with qubits placed in a single row~\cite{zindorf2024efficient,balauca2022efficient,zindorf2025multi}, see Ref.~\cite{ToffoliMolmer} for a proposal of implementing such operations natively in trapped ions.
Nevertheless, for contemporary prototypes of quantum computers, the overhead in DB-QITE can still pose challenges.
Instead, we will show that DB-DOI can be effective on noisy hardware by having $\D$ involve only single-qubit terms.
Additionally, as a rule of thumb for noisy quantum hardware, the first  step of DB-QITE roughly involves 3 queries to Hamiltonian simulation, while DB-DOI involves only one query.
For this reason, DB-DOI is preferred over DB-QITE when implementing the DBQA approach on quantum hardware.

On the other hand, we will see that a single DB-QITE step tends to reduce the energy by a larger amount than DB-DOI, which instead requires additional iterations to achieve a comparable energy reduction.
For both DBQAs we present numerical simulations for up to $L=20$ qubits.

Finally, we remark that we can use the quantum dynamic programming~\cite{QDP} approach for executing the recursion~\eqref{db-dbqite} with polynomial circuit depth but with an exponential circuit width as a trade-off. 
While the same can be done for DB-DOI, it would require coding the diagonal operator $\D$ in a quantum state.

\subsection{Warm-starts for DB-DOI} \label{sec:tailoring_vqe}
Ref.~\cite{Gluza_2024} introduced DB-DOI for eigenstate preparation, building on known convergence results~\cite{helmke_moore_optimization,smith1993geometric}, but it did not address how to specifically target ground states. In the present work, we demonstrate that warm-starts significantly enhance ground-state preparation, particularly when low gate counts are a priority. In the subsequent Ref.~\cite{gluza2024double}, DB-QITE was developed and also leverages warm-starts. We now discuss how warm-starts function in each of these methods, highlighting both the practical benefits observed here and directions for future optimization.

As we explained above, the sorting property of Brockett's DBF implies that as long as the initial state DB-DOI is the ground state of $\D$, then DB-DOI will rotate it to the ground state of $\h_0$.
To estimate the number of steps required, we qualitatively consider two key factors: the spectral gap of the Hamiltonian and the exponential convergence behavior of DBQAs. Once DB-DOI has advanced sufficiently, the operator $\h_k$ will be largely diagonal, and consequently the bracket $[\D, \h_k]$ becomes small. By “small”, we mean that the resulting unitaries $\hat R_k$ cannot change the energy by more than the spectral gap in a single step. Consequently, although the algorithm converges at the exponential rate characteristic of approximate gradient descent, the limited energy change per step naturally drives the state toward the closest eigenstate in energy. When DB-DOI is initialized with a warm start that is sufficiently close to the ground state, then the basin of attraction will indeed contain the ground state.

The concept of a warm start for DB-QITE arises naturally. Consider a low-energy state $\ket{\psi_0} = \hat U_0 \ket{0}$, where $\hat U_0$ prepares the initial state from the reference state $\ket{0}$. In this case, one can take $\hat U_0 \neq \id$ as the starting point, and the DB-QITE recursion then proceeds as
\begin{align}
    \hat U_{k+1} = e^{i\sqrt{s}\h}{\hat U_{k}}^\dagger e^{-i\sqrt{s}\ket{0}\bra{0} }\hat U_k e^{-i\sqrt{s}\h} \hat U_k\ .
    \label{db-dbqite2}
\end{align}
Ref.~\cite{gluza2024double} proved that in each step, the energy is improved.
This is essentially the same as Eq.~\eqref{db-dbqite} but with a different basic reflection operation, and it is as if we performed some $\hat U_0$ in place of the first DB-QITE step.

We next discuss warm-starting in DB-DOI, where we will work in the Heisenberg picture instead of the Schrodinger picture in the case of DB-QITE.
Set
\begin{align}
\h_\text{ws} = \hat U_0^\dagger \h_0 \hat U_0\  
\label{vqaxdbqa}
\end{align}
and for $k>1$ we consider the rotated Hamiltonians in Eq.~\eqref{h DOI} where $\hat U_k$ is defined by Eq.~\eqref{db-doi} rooted in the warm-start $\hat U_0$.
As we will see in Sec.~\ref{sec:unfolding}, 
one does not need to construct $U_0$ to have specific diagonalization properties.
Instead, for DB-DOI, the warm-starting mechanism can be effective by relying on methods that yield circuits $\hat U_0$ for approximating the ground state of $\h_0$.

We apply the DB-DOI unitary operations to the reference state $\ket{\psi_k}:= \hat U_k\ket 0$. The,n we work in the Heisenberg picture and recover the correct energy by setting
\begin{align}
    \tilde E_k:= \bra{\psi_k}\h_0\ket{\psi_k} = \braket{0|\h_{k}|0}\ .
\end{align}
The latter value converges to an eigenvalue by virtue of the diagonalization property of Brockett’s DBF, and consequently $\ket{\psi_k}$ serves as an approximation to the corresponding eigenstate that can be prepared on a quantum computer. In practice, the initial steps of the optimization are the most significant. We now turn to the explicit form of the DB-DOI unitaries.
\subsection{Explicit unfolding of DB-DOI with warm-start }
\label{sec:unfolding}
To summarize the above discussions, let us track a few steps of DB-DOI after a warm start.
For the first update of the Hamiltonian, we use Eq.~\eqref{vqaxdbqa}
so that the energy expectation value agrees with the loss function of the warm-start state $\ket{\psi_0}=U_0\ket \bfzero$ as follows:
\begin{align}
    \bra \bfzero \h_\text{ws}\ket \bfzero =   \bra{\psi_0 } \h_0\ket{\psi_0 }\ .
\end{align}
This exemplifies the difference between the Heisenberg and Schr\"odinger pictures: The left-hand side of the equality is in the former and the right-hand side is in the latter.

Recall that in the warm-starting convention we set $\hat U_0 \neq \id$.
In the second step we will set $r = \sqrt{s}$ and
\begin{align}
    \hat U_1 = \hat U_0 e^{ir \D} \hat U_0^\dagger e^{ir \h_0} \hat U_0 e^{-ir \D}\ .
    \label{eq:U1DBDOi}
\end{align}
With this we find $\h_1 =  \hat U_1^\dagger\h_0 \hat U_1$. This means that the state will be
\begin{align}
    \bra \bfzero \h_1\ket \bfzero =   \bra{\psi_1 } \h_0\ket{\psi_1 } 
\end{align}
where $\ket{\psi_1 } =\hat U_1\ket \bfzero$.
Notice again how in DB-DOI we work in the Heisenberg picture, and the reference state does not change, but rather the Hamiltonian is being rotated.
We could also rewrite $\hat U_1$ in Eq.~\eqref{eq:U1DBDOi} as
\begin{align}
    \hat U_1 = \hat U_0 e^{ir \D}   e^{ir \h_\text{ws}}   e^{-ir \D}\ ,
    \label{eq:U1ws}
\end{align}
but it would be only useful if it were easier to execute evolutions $e^{ir \h_\text{ws}}$ than the composed evolution $\hat U_0^\dagger e^{ir \h_0} \hat U_0$.

It is important to notice that $\hat U_0$ in Eq.~\eqref{eq:U1DBDOi} acts after the group commutator approximation, \emph{i.e.} the DB-DOI step acts before the warm-start.
This suggests that the DB-DOI step has the role of entangling the computational basis state subtly and preparing it for the action of the warm-start unitary.
However, note also that
\begin{align}
e^{-ir \D}\ket \bfzero = \ket \bfzero
\end{align}
whenever $\D$ is traceless and otherwise we acquire an immaterial global phase.
Thus, more explicitly, we have
\begin{align}
\ket{\psi_1 } &= \hat U_0  e^{ir \D} \hat U_0^\dagger e^{ir \h_0} \hat U_0\ket \bfzero = \hat U_0  e^{ir \D} \hat U_0^\dagger e^{ir \h_0} \ket{\psi_0 }
\end{align}
which means that the Hamiltonian evolution in the first step can indeed be interpreted as acting on the warm-start state $\ket{\psi_0}$.
We remark that it is not apparent a priori that such sequencing of unitary operations would lower the energy of the state -- the DBQA framework provides a sound theoretical grounding for that.

To advance by one more DB-DOI step, we set 
\begin{widetext}
\begin{align}
   \hat U_2 
   =& \hat U_1e^{ir \D} \hat U_1^\dagger e^{ir \h_0} \hat U_1 e^{-ir \D} \\
   =& 
\hat U_0 e^{ir \D} \hat U_0^\dagger e^{ir \h_0} \hat U_0 
e^{ir \D}\hat U_0^\dagger e^{-ir \h_0}\hat U_0 e^{-ir \D}\hat U_0^\dagger 
   e^{ir \h_0} 
\hat U_0 e^{ir \D} \hat U_0^\dagger e^{ir \h_0} \hat U_0 e^{-i2r \D}
   \label{equnfoldedG1}
\end{align}
\end{widetext}
Here, the point is that there are in total 4 queries to the evolution governed by the input Hamiltonian, 9 queries to the warm start, and 5 queries to evolutions governed by $\D$.
The latter count is lowered by fusing unitaries of the same type, \emph{e.g.} obtaining the trailing unitary $e^{-i2r \D}$. 
This leads to
\begin{align}
    \ket{\psi_2 } = \hat U_2\ket \bfzero
\end{align}
and again by disregarding a global phase, the implementation of the trailing diagonal unitary $e^{-i2r \D}$ can be omitted in $\hat U_2$.
We will say that Eq.~\eqref{equnfoldedG1} is the unfolded form of the second DB-DOI step.
To execute multiple steps of DB-DOI on quantum hardware, we first repeat such unfolding of the recursion for further steps until all unitaries appearing in the sequence are generated either by $\D$ or $\h_0$.

\subsection{Higher-order product formulas in DBQAs}
An essential step of the DBQA formalism is to specify an explicit strategy for compiling the  unitaries $\RDBI_k$ in Eq.~\eqref{eq:RkDBI} as quantum circuits. 
The most basic approach is to use the first-order group commutator formula in Eq.~\eqref{basic gc eq}.
However, we can replace $\RDBI_k$ by a higher-order approximation.
For example using Ref.~\cite[Eq.~(8)]{HOPFGC3} we can use invariance as was done in Eq.~\eqref{eq dropping eith} and  obtain the following higher-order DB-DOI recursion
\begin{align}
    	\hat U_{k+1} = \hat U_{k}
	e^{i r\phi\D}
	e^{ ir\h_k}
	e^{-ir(\phi+1)\D}
	e^{-i r(1-\phi) \h_k}
	e^{ir\D},\
	 \label{eqHOPF}
\end{align}
with $\phi = \frac  12 (\sqrt 5 -1)$ and $r = \sqrt{s}$.
This gives $\hat U_{k+1}^\dagger \h_0 \hat U_{k+1} \approx e^{s_k[\D,\h_k]}\hat U_{k}^\dagger\h_0\hat U_{k}e^{-s_k[\D,\h_k]}$,
with an error bounded as  $O(s_k^{2})$, and this could be beneficial when the exponential of the commutator can give a large energy gain and therefore should be approximated accurately.
In that case, other group commutator approximations could be considered~\cite{product_formula2013,kuperberg2023breaking,elkasapy2015length,elkasapy2016new,peetz2025hamiltonian}.
Whenever we use a product formula we will use  the identity $e^{i r_{k} \h_k} = \hat{U}_k^{\dagger}  e^{i r_k \hat{H}_0} \hat{U}_k$ and call Hamiltonian simulation~\cite{ChildsSu,Campbell2019random} for each appearance of an evolution operator in $\hat U_k$.

This recursive unfolding at each step $k$ results in a circuit depth that grows exponentially with $k$. More precisely, if a product formula involves $n$ evolution operations, then $k$ steps require $(2n+1)^k$ queries to Hamiltonian simulation. For instance, in Eq.~\eqref{eqHOPF} we have $n=2$, so DB-DOI entails $5^k$ queries, whereas the minimal first-order formula in Eq.~\eqref{db-doi} requires only $3^k$ queries. Although the latter demands fewer queries, it may yield smaller energy improvements. At present, we lack evidence that one approach is unambiguously superior. On very noisy hardware, the minimal first-order formula in Eq.~\eqref{db-doi} is preferable, whereas in other settings it may be advantageous to employ the higher-order formula of Eq.~\eqref{eqHOPF}, trading fewer recursion steps for potentially larger energy gains per step.

\begin{figure*}[ht]
  \centering
    \includegraphics[width=0.49\linewidth]{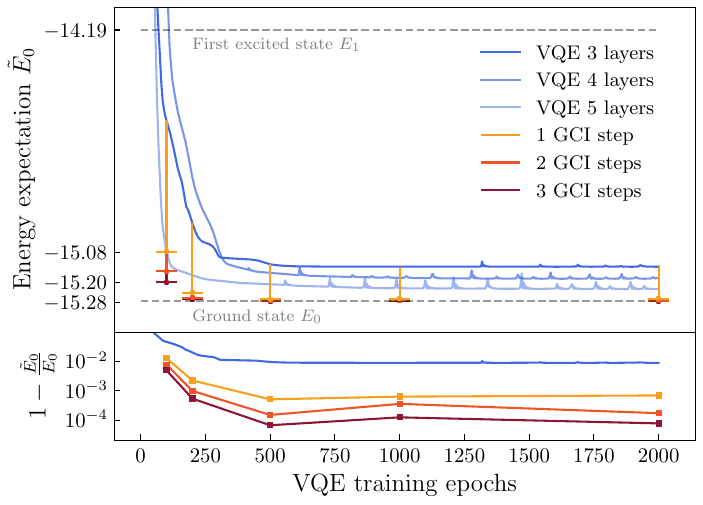} 	\includegraphics[width=0.48\linewidth]{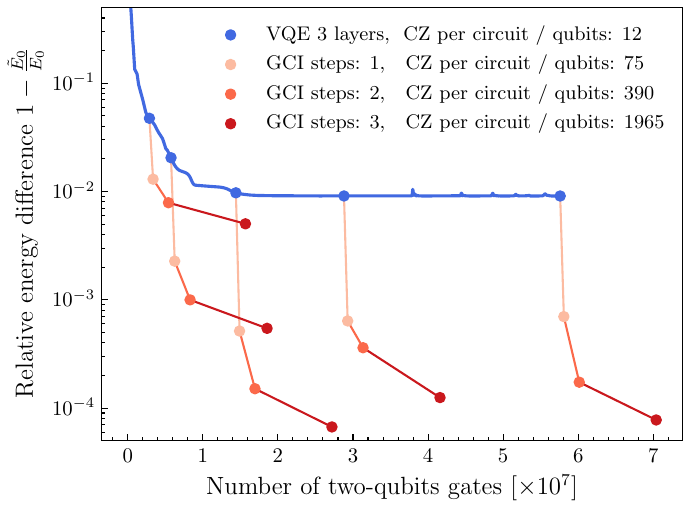} %
	\caption{Visualization of the 
	impact of DB-DOI~on cost function for a single VQE random 
	seed, see Tab.~\ref{tab:xxz_results} for statistical analysis. \emph{(left)}~Training of VQE (blue lines) 
	for 3, 4, and 5 layers of a Hamming-weight-preserving ansatz (hues of blue) achieved ground state energy residue $\Delta E\approx  1\%$ within $500$ training epochs.
	For more epochs, the initially rapid decrease in the cost function saturates and shows marginal improvement afterward. 
	We initialize DBQA with VQE for selected epochs $\in \{100, 200, 500, 1000, 2000\}$ and  optimize DBQA parameters with CMA-ES~\cite{cma}.
    In the bottom panel, we show the relative difference value between the achieved energy $\tilde E_0$ and the true ground state energy $E_0$.
	\emph{(right)} Token cost estimates of DB-DOI~by counting the total number of CZ gates required for the complete protocol, which includes the training (using the parameter-shift rule) of the VQE until a target epoch and the optimization of DBQA. The depth of the individual circuits are instead reported in the legend.}
    \label{fig:hw_xxz}
  \end{figure*}

\begin{table*}[ht]
\footnotesize
\begin{tabular}{c|cc|cc|cc|c|cc}
\hline \hline
\textbf{Layers} & \multicolumn{2}{c|}{\textbf{Warm-start}} & \multicolumn{2}{c|}{\textbf{1 DB-DOI step}} & \multicolumn{2}{c|}{\textbf{2 DB-DOI steps}} & \textbf{1 DBI step} &\multicolumn{2}{c}{\textbf{Long VQE training}} \\
\hline
 & \multicolumn{2}{c|}{$1-\tilde{E}_0/E_0$} & \multicolumn{2}{c|}{$1-\tilde{E}_0/E_0$} & \multicolumn{2}{c|}{$1-\tilde{E}_0/E_0$} & $1-\tilde{E}_0/E_0$ &\multicolumn{2}{c}{$1-\tilde{E}_0/E_0$} \\ 
\hline
 $3$ & \multicolumn{2}{c|}{$0.012\pm0.004$} & \multicolumn{2}{c|}{$0.0011\pm0.0007$} & \multicolumn{2}{c|}{$0.0005\pm0.0004$} & $0.0009\pm0.0006$ &\multicolumn{2}{c}{$0.010\pm 0.004$ } \\
 $4$ & \multicolumn{2}{c|}{$0.008\pm0.004$} & \multicolumn{2}{c|}{$0.0006\pm0.0005$} & \multicolumn{2}{c|}{$0.0002\pm0.0002$} & $0.0005\pm0.0004$ &\multicolumn{2}{c}{$0.004\pm0.002$} \\
 $5$ & \multicolumn{2}{c|}{$0.005\pm0.003$} & \multicolumn{2}{c|}{$0.0003\pm0.0002$} & \multicolumn{2}{c|}{$0.0001\pm0.0001$} & $0.0002\pm0.0002$ &\multicolumn{2}{c}{$0.003\pm0.002$} \\
\hline 
 &  Depth & Cumulative cost &  Depth & Cumulative cost &  Depth & Cumulative cost & - & Depth & Cumulative cost \\
\hline
$3$ & $12$ & $1.44\times 10^{7}$ & $75$ & $1.49\times 10^{7}$ & $390$ & $1.69\times 10^{7}$ & - &  $12$ & $5.76\times 10^{7}$  \\
$4$   & $16$ & $2.56\times 10^{7}$ & $95$ & $2.62\times 10^{7}$ & $490$ & $2.88\times 10^{7}$ & - & $16$ & $10.24\times 10^{7}$  \\
$5$  & $20$ & $4.0\times 10^{7}$ & $115$ & $4.07\times 10^{7}$ & $590$ & $4.38\times 10^{7}$ & - &  $20$ & $16.0\times 10^{7}$ \\
\hline
\hline
\end{tabular}
\caption{\label{tab:xxz_results} \textit{(above)} Relative energy difference for the $\XXZ$~ model of Eq.~\eqref{eq:XXZ_H} with $L=10$ qubits, and $\Delta = 0.5$. The estimates with their uncertainties were calculated using the median and the median absolute deviation of a sample of results obtained by repeating the execution $50$ times with different initial conditions.
\textit{(below)} Circuit depth expressed as the number of CZ gates per qubit, alongside the cumulative number of CZ gates used to reach $\tilde{E}_0$ (See App.~\ref{sec:training_setup}). 
Warm-started VQE approximations ($500$ epochs of training) are presented alongside DB-DOI results, executed considering both compiled (DB-DOI) and theoretical (DBI) approaches. 
For DBI, we compute $\RDBI_k$ through dense matrix representation, so there is no gate count. Longer VQE training (2000 epochs) is reported in the last column of the table.}
\end{table*}

\subsection{Variationally optimizing steps of DB-DOI }

We next notice that in each step $k$ of DB-DOI, we may select a different step duration $s_k$.
To study this further, we define for each $k$
\begin{align}
    \h_k(s) = e^{s [\D,\h_k]} \h_k(0) e^{-s [\D,\h_k]}
    \label{DBR}
\end{align}
as a function of a variable $s$
and call it a double-bracket rotation since it satisfies a Heisenberg equation involving two, not one, brackets
\begin{align}
    \partial_s\h_k(s) = [  [\D,\h_k(0)], \h_k(s)]\ .
    \label{eq:nonlinearHeisenberg}
\end{align}
This allows us to define a greedy optimization scheme of a cost function $f: \mathbb C^{2^L\times 2^L} \rightarrow \mathbb R_\ge$ by considering a global minimum of how much a double-bracket rotation can reduce the cost function
\begin{align}
	s_k = \text{argmin}_{s\in\mathbb R} f(\h_k(s))\ .
\end{align}
Given this optimizer we then set $\h_{k+1} = \h_k(s_k)$.

Moreover, we have the freedom to optimize in every step the operator $\D$.
For each DB-DOI step, we parametrize $\D_k$ as a classical Ising model with nearest-neighbor (NN) interactions:
\begin{equation}
\D_k(B^{(k)},J^{(k)})=\sum_{i=1}^L(B_i^{(k)}\Z_i + J_i^{(k)}\Z_{i+1}\Z_i),
\label{NNIsing}
\end{equation}
where $L$ is the number of sites in the chain, the parametrization is implemented through the coefficients $\{B_i\}_{i=1}^L$ and $\{J_i\}_{i=1}^L$, and the superscript $k$ highlights the optimization procedure is repeated for each DBQA step (see App.~\ref{sec:training_setup}). 
We refer the reader to Ref.~\cite{xiaoyue_strategy} for a detailed discussion on strategies for optimizing the $D_k$ operators.
As we will mention also explicitly below, we can achieve sizable energy gains with DB-DOI involving the constant magnetic field $\D(1,0)\equiv \sum_{i=1}^L\Z_i$.
However, for that step duration, the optimization is still essential.
\begin{figure}
\tikzset{blue dot/.style={fill=blue}}
\begin{quantikz}[row sep = 0.5, column sep = 8]
	\lstick{$\ket{1}$} & \ctrl{1}	& \qw & 
	\control[style=mediumpersianblue!70,"\theta"{mediumpersianblue, above left, xshift=-0.2, yshift=0.2}]{} & 
	\ctrl{2} & \qw & 
	\control[style=mediumpersianblue!70,"\theta"{mediumpersianblue, above left, xshift=-0.2, yshift=0.2}]{} 
	& \qw & \qw  & \dots & & \meter{} \\
	\lstick{$\ket{1}$} & 
	\control[style=mediumpersianblue!70,"\theta"{mediumpersianblue, above left, xshift=-0.2, yshift=0.2}]{}& 
	\ctrl{1} & \qw	& \qw  & \ctrl{2} & \qw & 
	\control[style=mediumpersianblue!70,"\theta"{mediumpersianblue, above left, xshift=-0.2, yshift=0.2}]{} 
	&\qw & \dots & & \meter{} \\
	\lstick{$\ket{0}$} & \ctrl{1} 	& 
	\control[style=mediumpersianblue!70,"\theta"{mediumpersianblue, above left, xshift=-0.2, yshift=0.2}]{}
	& \qw & 
	\control[style=mediumpersianblue!70,"\theta"{mediumpersianblue, above left, xshift=-0.2, yshift=0.2}]{} & 
	\qw & \ctrl{-2} & \qw & \qw & \dots & & \meter{} \\
	\lstick{$\ket{0}$} & 
	\control[style=mediumpersianblue!70,"\theta"{mediumpersianblue, above left, xshift=-0.2, yshift=0.2}]{}& 
	\qw		& \ctrl{-3}  & \qw  & 
	\control[style=mediumpersianblue!70,"\theta"{mediumpersianblue, above left, xshift=-0.2, yshift=0.2}]{} & 
	\qw & \ctrl{-2}& \qw & \dots & & \meter{}
\end{quantikz}
\vspace{-0.2cm}
\begin{center}
    $\underbrace{\hspace{4.8cm}}_{\text{1 layer}} \qquad \qquad \quad$
\end{center}
\caption{Hamming-weight-preserving architecture for $L=4$ qubits and $S = 2$. 
A single circuit layer consists of Reconfigurable Beam Splitter (RBS) gates (see Eq.~\eqref{eq:rbs}) connecting nearest- and next-nearest neighbors. 
We account for periodic boundary conditions by adding RBS gates connecting the last and first qubits.
After a chosen number of layers, we add measurements in the computational basis.
}
\label{fig:hwp}
\end{figure}

\vspace{1.5mm}

\vspace{1.5mm}
\section{Numerical results for \textup{XXZ}} Consider the one-dimensional XXZ Heisenberg model,
\begin{align}
  \h_{\XXZ} = \sum_{i=1}^{L} (\X_i \X_{i+1} +\Y_i\Y_{i+1} +\Delta \Z_i \Z_{i+1}) \, ,
  \label{eq:XXZ_H}
\end{align}
where the subscript $i$ indicates that the Pauli operators act on the $i$-th qubit, and $\Delta$ is the anisotropy parameter that decides whether the system is gapped or gapless in the thermodynamic limit.
The XXZ model \eqref{eq:XXZ_H} is well-understood by means of the Bethe ansatz~\cite{Levkovich_Maslyuk_2016, Sopena_2022, Ruiz_2024} and tensor networks~\cite{Murg_2012}.
In Sec.~\ref{section VQE}, we study DB-DOI with warm start by VQE~\cite{nepomechie2021betheansatzquantumcomputer} and use periodic boundary conditions (note the summation until $L$ in Eq.~\eqref{eq:XXZ_H}, and we set $X_{L+1}\equiv X_1$ etc.).
On hardware, it is often more natural to consider open boundary conditions ($X_{L+1}\equiv 0$ etc.), which we shall also use when studying larger system sizes.

Our objective is to achieve high fidelity for the ground-state preparation task, which we define qualitatively in light of the following considerations.
For quantum many-body systems, perturbation theory is inherently unstable, as typical vectors in a high-dimensional Hilbert space exhibit exponentially small overlaps; this phenomenon is sometimes referred to as the orthogonality catastrophe~\cite{PhysRevLett.124.110601}.
Consequently, fidelity with the ground state should be regarded as an especially sensitive measure of success.
In practice, physics often addresses this sensitivity by shifting from strictly quantitative to more qualitative criteria: instead of demanding high fidelity, one may seek a state of sufficiently low energy, and instead of requiring exact ground-state preparation, one may settle for a low-temperature state.
In this work, we are guided by the notion of ground-state fidelity witnesses~\cite{cramer2010efficient}, and we refer to a state $\ket{\psi}$ as a high-fidelity ground-state preparation whenever its energy expectation value $\langle \h_0\rangle_\psi$ lies substantially below the first excited-state energy $E_1$.
When this is true, then the preparation has energy close to the ground state energy $E_0$, and so the fidelity witness $\mathcal{F}$ defined as
\begin{align}
    \mathcal F \coloneqq 1 - (\langle \h_0\rangle_\psi -E_0)/ (E_1-E_0)
\end{align}
will  have a non-trivial value $\mathcal F\ge 0$.
Given that $s\mathcal F$ is a lower bound to the fidelity of the preparation, we see intuitively that reaching energy falling within the spectral gap of the model $E_0 \le \langle \h_0\rangle_\psi \le E_1 $ implies a non-trivial fidelity to the ground state.
We will say that a preparation reached ground state with high fidelity when the energy expectation value is below the middle of the spectral gap
\begin{align}
    \langle \h_0\rangle_\psi \le (E_0+E_1)/2\ .
    \label{high fidelity}
\end{align}
The choice of setting it in the middle is arbitrary, but it is both permissive enough given the orthogonality catastrophe and stringent enough because all such states will have the actual fidelity $F\ge 50\%$.
Below, we will see that warm starts by parametrized circuits can bring the energy expectation value to within the spectral gap.
On the other hand, reaching a high-fidelity regime in this qualitative sense becomes difficult for about a dozen qubits.
Fortunately, DB-DOI can amplify the fidelity of its warm-start.
Moreover, we will also consider a problem-specific variational ansatz that performs better. With this, we show that DB-DOI can consistently improve it to arrive in the high-fidelity regime as defined by Eq.~\eqref{high fidelity}.

\subsection{Exploring numerically-simulated variational optimization of parametrized circuits}
\label{section VQE}
The variational quantum eigensolver (VQE)~\cite{Peruzzo_2014} is a variational quantum computing routine that minimizes the same cost function by optimizing a parametric quantum circuit $\vqe$, \emph{i.e.} varies parameters $\bm{\theta}$ to find an optimal set of parameters 
\begin{equation}
\bm{\theta}^* = \text{argmin}_{\bm{\theta}} \bigl\{ \braket{0|\vqe^{\dagger} \h_0  \vqe|0} \bigr\}\ 
\label{vqe_argmin}
\end{equation}
that minimizes the energy. 
Although  VQE is known for its variational character,  trainability and expressibility problems appear~\cite{McClean_2018, Larocca2024, Holmes_2022, cerezo2021variationalReview, anschuetz2022beyond, cerezo2023does}, limiting its effectiveness. 
However, for DBQAs we only need an approximation of the ground state and VQE might be effective for this despite its limitations. 
Thus we will use VQE as a warm start for DBQA and set $\hat U_0 = \vqes$ in Eq.~\eqref{vqaxdbqa}.

As a remark, in this work, we do not use the variational aspect as an optimization loop involving a quantum computer.
Since we only need a sufficiently good initial state, we will instead numerically simulate the VQE training process. The optimal set of parameters $\bm{\theta}^*$ obtained from simulation can then be used to initialize the parametrized quantum circuit on quantum hardware as a warm-start for DB-DOI.
We quantitatively assess whether this approach is viable for DB-DOI.
As a general remark, even for $L\gg 10$ qubits where VQE training becomes ineffective, DB-DOI could benefit from a VQE warm-start by subdividing $L=K L'$ into $K$ blocks and then initializing DB-DOI from a product state of $K$ VQEs computed by a classical variational loop which we will present below for only 1 block.

We will report below quantitative results for preparing the ground state of the $\XXZ$~ model, which satisfies a total-spin conservation~\cite{PhysRevLett.113.127204}.
For an even number of qubits $L$ --- as in the examples examined in this work ---  and total spin $S$, the ground state of the $\XXZ$~ Hamiltonian is constrained to 
the half-filling subspace, \emph{i.e.} a superposition of states associated to total spin $S = L / 2$. 
We account for this symmetry by using a \emph{Hamming-weight-preserving} VQE ansatz \cite{RAISSI2019686, Anselmetti2021, Arrazola2022, Johri2021, Kerenidis2022, Jain2024,representationtheorygeometricquantum, monbroussou2023trainability, symmetrybreakinggeometricquantum, Larocca2024,cerezo2023does,raveh2024estimatingbetherootsvqe,farias2024}. This reduces the search space to a Hilbert space of size $\binom{L}{L/2}$, providing a polynomial compression.
Even though this compression is not ideal, \emph{i.e.} exponential, fulfilling the system symmetries allows for faster training due to circuit updates being constrained to a subspace of interest \cite{monbroussou2023trainability, Cherrat_2024,crognaletti2024, PhysRevC.106.034325, PhysRevE.107.024113}. 
One way to construct a Hamming-weight-preserving ansatz consists of creating a network of Reconfigurable Beam-Splitter gates 
\begin{center}
\begin{quantikz}[row sep = 10]
	& \ctrl{1} & \qw \\
	& \control[style=mediumpersianblue!70,"\theta"{mediumpersianblue, above left, xshift=-0.2, yshift=0.2}]{} & \qw \\
\end{quantikz}
$
\quad = \quad \begin{bmatrix} 
1 & 0 & 0 & 0 \\ 
0 & \cos \theta & \sin \theta & 0 \\
0 & -\sin \theta & \cos \theta & 0 \\
0 & 0 & 0 & 1 \\
\end{bmatrix}.~\quad\refstepcounter{equation}(\theequation)\label{eq:rbs}
$
\end{center}
Such a gate is a common building block for constructing circuits that preserve the Hamming weight of bitstrings that label computational basis states in a given superposition. 
In Fig.~\ref{fig:hwp} we describe the circuit architecture implemented in our numerical experiments.
For completeness, we also report in App.~\ref{sec:app_vqe} results obtained with a hardware-efficient ansatz~\cite{Kandala_2017}, which explores the full Hilbert space of dimension $2^{L}$. The enhanced expressibility increases the difficulty of training, yet combining with DBQA once again proves beneficial. While employing a symmetry-aware ansatz improves convergence, it does not directly influence the performance of DBQA itself when the fidelity of the warm start is held fixed.
\renewcommand{\arraystretch}{1.5}

\subsubsection{Results}
After training the VQE circuit using gradient descent with the parameter-shift rule \cite{Schuld_2019} and the ADAM optimizer \cite{kingma2017adam} until a target epoch, we use it as a warm start for DB-DOI.
For each DB-DOI step, we parametrize $\D_k$ as the classical Ising model with nearest-neighbor (NN) interactions as in Eq.~\eqref{NNIsing} and optimize the parameters using the CMA genetic algorithm \cite{cma}, which does not require explicit computation of the gradients. 
This parametrization allows us to compile $e^{-it\D_k}$ with at most two layers of $\mathrm{CZ}$ gates, and we use two steps of first order Trotter-Suzuki decomposition~\cite{ChildsSu} for a short-depth compilation of $e^{-it\h_{\XXZ}}$, see App.~\ref{sec:xxz_compiling}.
A detailed description of the procedure by which we obtained the results we present in this section can be found in App.~\ref{sec:training_setup}.

To visualize the advantages of DB-DOI with VQE warm-starts, in Fig.~\ref{fig:hw_xxz} we show the results for $L=10$ qubits obtained by executing the procedure once.
We find that DBQA halves the energy residue if applied to the early training epochs, and essentially reaches the ground state when executed later in the process. 
More specifically, if we use three layers of the ansatz from Fig.~\ref{fig:hwp}, so a warm-start circuit with depth of $12$ CZ gates per qubit, then following up with one DB-DOI step yields a circuit with depth $75$ CZ gates per qubit.
We quantify its performance through the energy approximation ratio 
\begin{align}
    \Delta E := (\tilde E_0 - E_0)/E_0\ ,
\end{align} 
where $\tilde E_0$ is the energy of DB-DOI~and $E_0$ is the true ground state energy.
Fig.~\ref{fig:hw_xxz} shows an improvement by an order of magnitude from $\Delta E\approx 1\%$ to $\Delta E\approx 0.1\%$.
This cost function gain is statistically significant, see Tab.~\ref{tab:xxz_results} where we consider $50$ executions of DB-DOI~for each VQE circuit configuration.
Consistently,  DBQA steps  improve energy estimation for all depths analyzed.
Energy measurements can be translated to fidelity lower bounds~\cite{cramer2010efficient} as described in App.~~\ref{sec:training_setup}. Tab~\ref{tab:xxz_hwpa_fid} shows that one step of DB-DOI~allowed us to get $\mathcal{F}\ge (99.6\pm 0.3)\%$ fidelity.

By increasing the circuit depth of the chosen ansatz of Fig.~\ref{fig:hwp} to $20$ CZ gates per qubit, it can be trained to reach~$\Delta E\approx0.4\%$. 
In general, whenever an initialization method is improved to match the previously top-performing DB-DOI~circuit, DBQA can be interfaced with that enhanced initialization.
Indeed, when initialized with the $20$ CZ gates per qubit circuit depth with $\Delta E \approx 0.4\%$, just one DBQA step again gives an order of magnitude gain $\Delta E \approx 0.03\%$.
We were not able to reach $\Delta E \approx 0.03\%$ by training deeper VQE circuits.

\begin{table}[ht]
\footnotesize
\begin{tabular}{c|c|c|c|c}
\hline \hline
\textbf{Layers} & \textbf{Warm-start} & \textbf{$k=1$} & \textbf{ $k=2$} & \textbf{$k=3$} \\
\hline
3 & $0.83\pm 0.06$ & $0.95\pm0.01$ & $0.993 \pm 0.006$ & $0.997\pm0.003$\\
4 & $0.89\pm 0.05$ & $0.992\pm0.007$ & $0.997 \pm 0.003$  & $0.998 \pm 0.001$\\
5 & $0.93\pm 0.04$ & $0.996\pm0.003$ & $0.998 \pm 0.002$ & $0.9992 \pm 0.0008$\\
\hline
\hline
\end{tabular}
\caption{\label{tab:xxz_hwpa_fid} Fidelity  lower bounds~\cite{cramer2010efficient} (see App.~\ref{sec:training_setup}) extending results presented in Tab.~\ref{tab:xxz_results}.  
}
\end{table}

The DB-DOI~circuits executed to produce Fig.~\ref{fig:hw_xxz} involve a VQE warm-start with $3$ layers of the ansatz from Fig.~\ref{fig:hwp}, which translates to circuit depth of $12$ CZ gates per qubit, and then one DB-DOI step yields depth $75$ while two steps have depth $390$.
The right plot in Fig.~\ref{fig:hw_xxz} shows the training cost of the algorithm quantified in terms of the cumulative number of CZ gates during optimization, which captures the runtime of the classical optimization. More details about the cumulative cost metric can be found in App.~\ref{sec:training_setup}. 
We find that the cost of training is dominated by VQE when running a few DBQA steps.
For tasks requiring high ground state preparation fidelity, VQE and DBQA should be used in sequence, as individually each would necessitate unnecessarily large token expenditures, \emph{i.e.} cost of execution on commercial hardware. 
This training cost reduction is summarized in Table~\ref{tab:xxz_results}, where we report the gate count in the case where DB-DOI~has been compiled into circuits via the DB-DOI formalism (see appendix~\ref{sec:xxz_compiling}).

Alongside compiled DB-DOI circuits, in the next-to-last column of Tab.~\ref{tab:xxz_results}, we show DB-DOI~results using DBI unitaries $e^{s_k[\D_k,\h_k]}$ which do not involve the product formula approximations and find that they perform similarly to DB-DOI, indirectly highlighting the usefulness of the higher-order approximation in Eq.~\eqref{eqHOPF}.

Further investigations are presented in Apps.~\ref{sec:app_scaling},~\ref{sec:app_vqe}~and~\ref{sec:j1j2}, where we consider \textit{i)} one-dimensional XXZ targets of sizes $L\in\{4,6,8,10,12\}$, \textit{ii)} a VQE architecture which does not respect XXZ symmetries and so requires longer training, but yields better warm-start energy in shorter depth, and \textit{iii)} next-nearest-neighbors interactions in the target Hamiltonian. 
In all these cases studied, we verify that DB-DOI remains advantageous.

\subsection{Exploring numerically-simulated Hamiltonian variational ansatz}

The Hamiltonian variational ansatz is well-suited for the XXZ model.
In particular, we will consider  up to $L=20$ qubits with open boundary conditions and $\Delta=1$, which leads to the gapless model XXX in the thermodynamic limit. To begin, we  use the simple observation that the XXX model is a sum of singlet projections and thus for XXX on a line, the tensor product of singlet states 
\begin{align}
    \ket{\text{Singlet}} = 2^{-L/4}(\ket{10}-\ket{01})^{\otimes L/2}
\end{align}
is in the ground state of half of the terms in the Hamiltonian.
We can further improve on it using the problem-specific Hamiltonian Variational Ansatz (HVA)~\cite{bosse_Heisenberg_2022, kattemolle_vanwezel_2022,wecker2015progress}
\begin{align}
\ket{\text{HVA}_p(\mathbf{t})} = \prod_{j=1}^{p}(e^{-it_{j,e} \h_e}e^{-it_{j,o} \h_o})\ket{\text{Singlet}}
\end{align}
where $\h_e, \h_o$ are obtained from $\h_0$ by restricting to summation over odd (even) indices $i$ such that all terms in $\h_e, \h_o$ commute. 
To understand this ansatz, notice that for $p=1$,
\begin{align}
\ket{\text{HVA}} = e^{-it_{0} \h_e}e^{-it_{1} \h_o}\ket{\text{Singlet}}= U_\text{HVA}\ket{\text{Singlet}} \, .
\end{align}
The first unitary acting on the singlets on the even sublattice has the action of creating singlets on the odd sublattice.
Thus the layers can be interpreted as creating a balance between the location of the singlets and putting the singlet sublattices into a superposition.
Ref.~\cite{gluza2024double} demonstrated that HVA works well as a warm-start for DB-QITE, and  below we will  assess its use for warm-starting DB-DOI.

\subsubsection{Results}

\begin{figure*}[ht]
  \centering
    \includegraphics[width=\linewidth ]{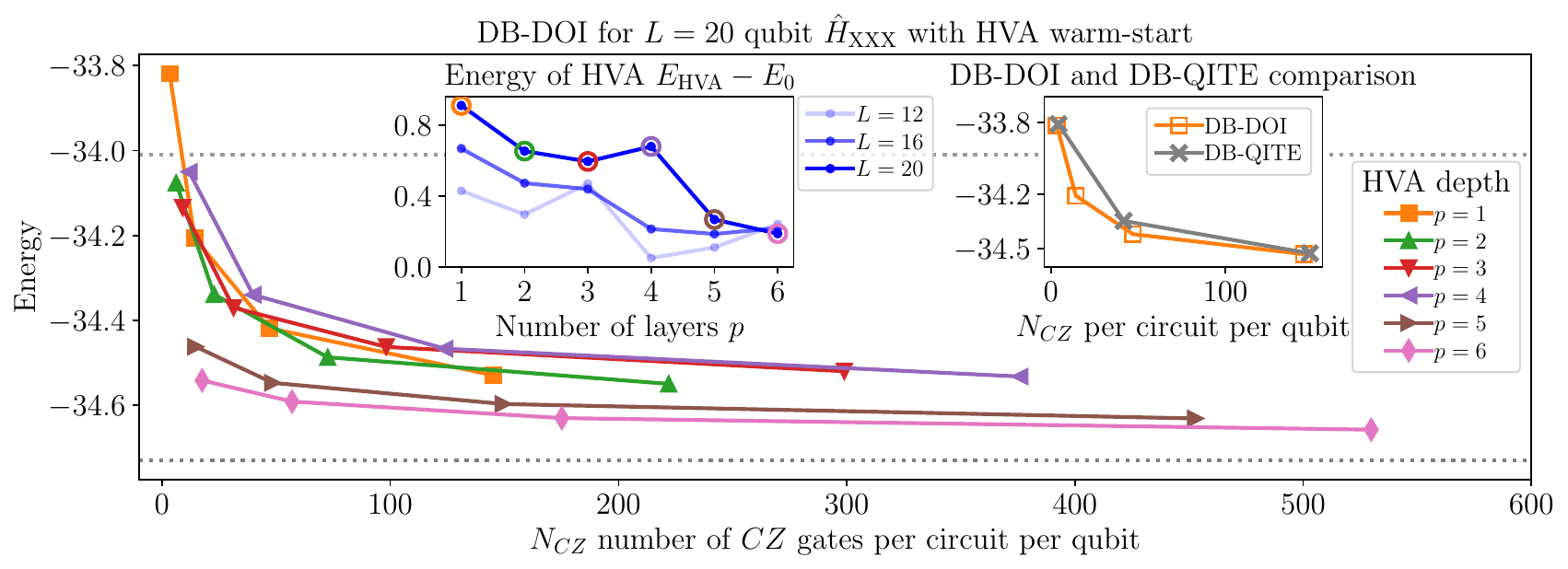} 	
	\caption{Performance of DB-DOI applied to the XXZ model with $\Delta = 1$ for $L=20$ qubits. The main panel shows the energy as a function of the number of CZ gates per circuit per qubit, with points corresponding to $k=0,1,2,3$ DB-DOI steps (different colors indicate different HVA ansatz depths $p$ for $k=0$). DB-DOI is warm-started from HVA states of varying depth (left inset), where the colored circles mark the DB-DOI initial energies in the main panel. 
    In the left inset, we show the relative energy to the ground-state energy $\lambda_0$ to visualize the system size scaling of HVA for $L=12, 16 20$ qubits.
    To explore HVA for varying system sizes $L=12,16,20$, we off-set the changing target energy by plotting the HVA energy relative to the exact ground state ($\Delta E_{\rm HVA} = E_{\rm HVA}-E_0$) as a function layer number $p$ which suggests that for $p=1$ HVA is limited by expressivity while for $p\ge 3$ it is possible to find improved ground state approximations but training becomes a limitation. The right inset compares DB-DOI and DB-QITE for the same initialization, showing that DB-DOI achieves similar energy reductions as DB-QITE with a few hundred CZ gates per qubit. Note that the number of CZ gates plotted refers to the circuit depth  and does not include the training's cost.}
    
    \label{fig:hva}
  \end{figure*}

We observe that the HVA with $p=1,2,\ldots,6$ can yield excellent approximations to the XXX ground state. 
When it performs well, it has the advantage of a relatively low gate count. 
However, it might be difficult to improve the quality of the warm start, as illustrated in the left inset of Fig.~\ref{fig:hva}: when training  variational circuits with the COBYLA optimizer\textemdash even with randomized basin hopping\textemdash  the objective function does not always decrease monotonically with increasing number of layers. This “hit-and-miss” behavior may reflect the difficulty of HVA training, similarly to the difficulties due to barren plateaus in VQEs~\cite{Park2024hamiltonian,WiersemaHVA,Larocca2022diagnosingbarren}. Furthermore, while the HVA provides an excellent approximation for the XXX ground state, its performance is not guaranteed for other systems. In such cases, or when further circuit improvement is needed, DBQAs offer a viable alternative. We find that for all values of $p$ considered, DB-DOI consistently improves upon the warm-start HVA energy, see main panel of Fig.~\ref{fig:hva}.

Finally, the right inset in Fig.~\ref{fig:hva} shows that DB-DOI achieves performance comparable to DB-QITE. DB-DOI has the advantage of a very inexpensive initial step relative to DB-QITE, though each step reduces the energy by a smaller amount. This limitation can be mitigated: starting from the same HVA warm-start with $p=1$, 
similar energies can be obtained by either DB-DOI with $k=3$ steps or DB-QITE with $k=2$ steps, both having comparable CZ gate counts.
It is important to note that the DB-QITE gate counts assume all-to-all connectivity; on fixed architectures without adaptive use of auxiliary qubits and measurements, the actual gate count would be higher than reported.

\section{ Execution of DB-DOI on IBM quantum hardware} \label{sec:ibmq}
Using remote access, we performed a proof-of-concept experiment on an IBM superconducting quantum chip for the $L=10$ qubit XXZ model with open boundary conditions. The initial state is obtained by training two layers of nearest-neighbors Hamming-weight-preserving gates on a classical simulator, acting of the alternating product state $|01\cdots 01\rangle$. The diagonal operator consists of single RZ rotations as well as RZZ interactions between next-neighbors, and the time evolution is implemented with a single step of a second order Trotter formula. 

Error mitigation is a crucial technique for leveraging the computational power of noisy quantum devices. Our protocol relies on a noise renormalization method, which has proven effective in prior works~\cite{teplitskiy2024,noise_estimation,kiss2024moments,kiss2025neutrino}. The central idea is to assume a depolarizing noise model, estimate its parameters directly on the quantum device, and then invert its effect during post-processing.

We recall that a  depolarizing noise model $\mathcal{N}_p[\cdot]$ characterized by a parameter $p$, acts on the expectation value of a Pauli observable $\sigma$ under a quantum state $\rho$ as
\begin{equation}
    \text{Tr}\left(\sigma \mathcal{N}_p[\rho]\right)= (1-p)  \text{Tr}\left(\sigma \rho\right).
\end{equation}
Provided the parameter $p$ is known, one can correct the noisy expectation values by dividing them by the factor $(1-p)$. Here, we can simply estimate this value by replacing the alternating state with the all-zero state $|0\cdots 0\rangle$, which is invariant under the VQE ansatz. The time evolution is mitigated by performing the first half of the product formula in the forward direction and the second one backward. 
This method is only effective under depolarizing noise, which is not always an accurate representation of real devices. However, the situation can be improved using Pauli twirling \cite{Pauli_Twirling}. 
Readout errors are mitigated by calibrating the device \cite{MEM}, as well as twirling the measurement channel \cite{TREX}.  

We provide the results for $L=10$ qubits, with and without the DB-DOI  step, in Table~\ref{tab:xxz_ibmq}.
To parametrize the error-mitigation, we used $50$ twirls with $1000$ shots each, placed on the $156$-qubit chip \texttt{ibm\_fez}.
We observe that error mitigation is particularly important when performing the DB-DOI iteration, which significantly increases the depth, as we can outperform the bare VQE. 

\begin{table}[ht]
\footnotesize
\begin{tabular}{c|c|c}
\hline \hline 
 & \textbf{Warm start} & \textbf{1 DB-DOI step}  \\
\hline
CZ depth  &4 &22 \\
\hline 
statevector &$ -14.01$&$-14.27 $\\
raw &$-13.14\pm0.01 $&$-10.60\pm0.01$ \\
mitigated&$-13.95\pm 0.01 $&$ -14.16\pm 0.04 $\\
\hline
\hline
\end{tabular}
\caption{\label{tab:xxz_ibmq} 
Energy computed on the superconducting quantum chip \texttt{ibm\_fez} with one DB-DOI step on top of a two-layer warm start. 
The error bars correspond to a $95\%$ confidence interval on the mean computed via Bayesian data augmentation \cite{kiss2024moments}. 
The target energy is $-14.36$. }
\end{table}

\section{Numerical emulation of DB-DOI for Quantinuum's quantum hardware}

Our experiment on IBM’s superconducting quantum chip suggests that error mitigation is required for DB-DOI to outperform VQE.
However, this requirement may not hold on other quantum platforms, whose distinct noise profiles, coherence times, and native gate fidelities can alter the relative performance of the quantum algorithms.

For instance, Table.~\ref{tab:xxz_quantinuum} contains the numerical simulation results executed on Quantinuum Qnexus, and the model \textit{`H1-Emulator`} reflects the noise characteristics of Quantinuum's H1 hardware.
Note that we used open boundary conditions for the execution on IBM quantum hardware in Sec.~\ref{sec:ibmq}, whereas here we use periodic boundary conditions. Consequently, the target energy is $-15.28$ compared to the previous $-14.36$. We make this choice to present a slightly different example that fits naturally with Quantinuum’s all-to-all qubit connectivity.

We found that one step of DB-DOI is predicted to improve on VQE, despite the expected hardware noise, and no further post-processing is required. 
To achieve that, it was important to adapt both the VQE warm-start and Hamiltonian simulations in subsequent DB-DOI steps to the native gates supported by the platform.
This reduces the physical runtime and hence exposure to decoherence. Additionally, when compiling with native ZZ-gates, 
any erroneous calibration of the ZZ-gate duration only results in a longer or shorter XXZ Hamiltonian simulation duration during the DB-DOI step; as in both cases they are related to the required cost-function Hamiltonian, the performance impact is effectively due to modifying the DB-DOI step duration $s$.
This is advantageous to the performance of both VQE and DBQA compared to using CZ gates, which require very precise timing of the underlying native ZZ-gates to achieve an accurate implementation of Hamiltonian simulation.

In addition, we find that although two DB-DOI steps produce a larger energy drop in the ideal (noise‐free) setting, in the presence of realistic device noise, the energy gain is substantially diminished. This is likely because the circuit depth of DB-DOI grew to be large enough to compound decoherence and accumulate gate errors.
This is thus an example of the break-even range when using state-of-the-art quantum hardware without quantum error-correction.

\renewcommand{\arraystretch}{1.5}
\begin{table}[ht]
\footnotesize
\begin{tabular}{c|c|c|c}
\hline \hline
\textbf{ } & \textbf{Warm-start} & \textbf{1 DB-DOI step} & \textbf{2 DB-DOI steps} \\
\hline
statevector & $-14.16$ & $-14.62$ & $-14.76$\\
H1-LE & $-14.19\pm0.16$ & $-14.63\pm0.09$ & $-14.75\pm0.14$\\
H1-Emulator & $-13.93\pm0.20$ & $-14.59\pm0.23$ & $-13.65\pm0.12$\\
\hline \hline
\end{tabular}
\caption{\label{tab:xxz_quantinuum}Simulation results on Quantinuum's H1 emulators showing the energy expectation of the open boundary Heisenberg XXZ model with $L=10$ and $\Delta=0.5$ after a 2-layer VQE warm-start followed by $k=1$ and $k=2$ DB-DOI steps. Both the VQE and hamiltonian simulation are adapted to the platform's supported native gates. The target energy is of -15.28.}
\end{table}

\vspace{1.5mm}
\section{Conclusions}\label{sec:conclusions} Current quantum hardware supports circuits with dozens to hundreds of $\mathrm{CZ}$ gates per qubit. Nevertheless, preparing high-fidelity ground-state approximations remains challenging due to the absence of effective compilation methods: VQEs suffer from training limitations in large architectures, while QPE requires fault tolerance. We suggest that the recently introduced DBQA framework~\cite{Gluza_2024} offers a promising route to compiling circuits that leverage the capabilities of near-term devices.

Our results indicate that combining VQE with DBQA yields lower energies with shorter training times compared to VQE alone. This hybrid approach may be particularly advantageous on early fault-tolerant architectures, when powerful quantum processors will be remotely accessible. Our implementation~\cite{repository} within \texttt{Qibo}~\cite{Efthymiou_2021} provides circuit compilations compatible with any QASM-based interface, making experimental demonstrations readily available.

A limitation of our approach is its reliance on sufficiently accurate initializations, which become harder to obtain as system size grows. However, Fig.~\ref{fig:hw_xxz} shows that DBQA markedly improves \emph{\quotes{undertrained}} initializations, reducing the energy even when applied for only one or two steps. Moreover, DBQA is agnostic to the initialization method and can be combined with diverse strategies; see Fig.~\ref{fig:hva} for DB-DOI warm-started by HVA on up to $20$ qubits, and App.~\ref{app:comparison} for additional comparisons. Related two-step approaches, such as quantum subspace expansion~\cite{QSE_science,Yoshioka_Krylov_25}, are also natural candidates for early fault-tolerant hardware and stand to benefit from warm-starts.

Although rigorous convergence guarantees for DB-DOI require circuit depths that grow exponentially with the number of steps, our numerics demonstrate substantial improvements with only a few steps, provided the initial state has energy below the first excited level. This makes DBQAs particularly suited to early fault-tolerant devices, where shallow circuits are essential but algorithmic guarantees remain valuable. Looking ahead, DBQAs may also serve as an effective initialization for QPE. In this context, quantum dynamic programming~\cite{QDP} could further reduce circuit depth, enabling DBQAs to act as a bridging tool between near-term and fault-tolerant eras.

Finally, it would be worthwhile to investigate how DBQAs could more broadly enhance variational optimization. An alternative direction to this work would be to use DBQA as a warm start for variational algorithms. Such an approach would connect naturally with a wide range of optimization techniques—Riemannian gradient flows~\cite{riemannianflowPhysRevA.107.062421,schulte2010gradient}, natural gradients~\cite{Stokes_2020}, and quantum imaginary-time evolution~\cite{mcArdle2019} among others—opening opportunities to combine their complementary strengths.

All results presented in this work are reproducible using the open-source 
code at Ref.~\cite{repository}.\\

\vspace{1.5mm}
\textbf{Acknowledgments.---}
We acknowledge the use of IBM Quantum services for this work. The views expressed are those of the authors and do not necessarily reflect the official policy or position of IBM, the IBM Quantum team or Quantinuum.
We thank K. Bharti, G. Crognaletti, A. Ghanesh, K. Mitarai,  A. Sopena, and K. Yamamoto for useful discussions. MR and OK are supported by the CERN Doctoral Program through the CERN Quantum Technology Initiative.
JS and NN are supported by the start-up grant for Nanyang Assistant Professorship of Nanyang Technological University, Singapore.
XL, KUG, and MG are supported by the Presidential Postdoctoral Fellowship of the Nanyang Technological University, Singapore.
JK gratefully acknowledges support from Dr. Max R\"{o}ssler, the Walter Haefner Foundation, and the ETH Z\"{u}rich Foundation. ZH acknowledges support from the Sandoz Family Foundation-Monique de Meuron program for Academic Promotion. 
This work was further supported by the National Research Foundation Singapore, under its Quantum Engineering Programme 2.0 (National Quantum Computing Hub, NRF2021-QEP2-02-P01). STG and JYK acknowledge funding support from A*STAR C230917003.

\bibliography{references.bib}
\clearpage
\newpage
\clearpage

    \newpage
    
\onecolumngrid
    \begin{center}
    \Large \textbf{Supplemental Material}
\end{center}

\tableofcontents

\section{\label{sec:app_scaling} Tackling different XXZ sizes}
In this section we consider one-dimensional XXZ models involving from $4$ to $12$ qubits. For each case analyzed,  we train a four-layer Hamming-weight-preserving ansatz implementing the architecture introduced in Fig.~\ref{fig:hwp} with an Adam optimizer for $300$ iterations. The optimization details are the same presented in detail in appendix~\ref{sec:training_setup}. We repeat the training $10$ times for each configuration, initializing the VQE circuit with a different set of angles sampled from the uniform distribution $\mathcal{U}(-\pi, \pi)$. 
We apply three steps of preconditioned DB-DOI loading each trained VQE at epochs $100$, $200$, and $300$. Each DB-DOI has been optimized through $50$ iterations of a CMA-ES~\cite{cma} algorithm (see appendix~\ref{sec:training_setup} for DB-DOI optimization details).

\renewcommand{\arraystretch}{1.5}
\begin{table*}[ht]
\footnotesize
\begin{tabular}{c|cc|cc|cc|cc}
\hline \hline
\textbf{Qubits} & \multicolumn{2}{c|}{\textbf{VQE training}} & \multicolumn{2}{c|}{\textbf{1 DB-DOI step}} & \multicolumn{2}{c|}{\textbf{2 DB-DOI steps}} &\multicolumn{2}{c}{\textbf{3 DB-DOI steps}} \\
\hline
 &  Depth & $\mathcal{F}$ &  Depth & $\mathcal{F}$ &  Depth & $\mathcal{F}$ & Depth & $\mathcal{F}$ \\ 
\hline
 $4$ & $16$ & $1$ & $95$ & $1$ & $490$ & $1$ & $2465$ & $1$ \\
 $6$ & $15$ & $1$ & $88$ & $1$ & $457$ & $1$ & $2298$ & $1$ \\
 $8$ & $16$ & $0.947 \pm 0.024$ & $95$ & $0.9977\pm0.0017$ & $490$ & $0.9990\pm0.0008$ & $2465$ & $0.9995\pm0.0005$ \\
 $10$ & $16$ & $0.889\pm0.016$ & $95$ & $0.9928\pm0.0012$ & $490$ & $0.9972\pm0.0012$ & $2465$ & $0.9987\pm0.0005$ \\
 $12$ & $15$ & $0.600\pm0.091$ & $92$ & $0.919\pm0.067$ & $473$ & $0.941\pm0.048$ & $2382$ & $0.956\pm0.037$ \\
\hline
\hline
\end{tabular}
\caption{\label{tab:scaling} Fidelity lower bound $\mathcal{F}$ (see Eq.~\eqref{eq:fid_approx}) for the XXZ model of Eq.~\eqref{eq:XXZ_H} with number of qubits $L
\in \{4,6,8,10,12\}$ and $\Delta=0.5$. Standalone VQE data are presented alongside DB-DOI~results, obtained through compiled DB-DOI circuits. We show the depth of the utilized circuit in the left sub-column for each experiment setup. The estimates with their uncertainties were calculated using the median and the median absolute deviation of a sample of results obtained
by repeating the execution ten times with different initial conditions.}
\end{table*}

The obtained results are shown in Tab.~\ref{tab:scaling}, and highlight how the  DB-DOI~approach starts to be impactful on the ground state preparation with $L \geq 8$ qubits. 
In fact, the application of DB-DOI after the VQE training allows us to reach a remarkable increase of the fidelity lower bounds even considering only one DB-DOI step. 
For $L<8$ qubits, the VQE can be used as standalone because it suffices to obtain the almost exact ground state.    
Alongside the fidelity lower bounds, we show the depth of the executed circuits computed as the total number of two-qubit gates per qubit. 
We observe that the depth of the circuits remains approximately constant.
This is due to how we compile the DB-DOI in terms of quantum gates.
In fact, the contributions due to VQE absorption in the compilation are fixed by the number of layers, and the size of the oracles implementing the Hamiltonian and diagonal evolutions are fixed by the number of time steps and the order of the respective Trotter-Suzuki decomposition.
Importantly, these results are obtained assuming each of the VQE circuits composed of four layers.
Increasing the size of the problem may require larger VQE circuits, or - in general - more expensive warm-start approximation unitaries.
In those cases, the depth of the required DB-DOI~circuit will suffer of the scalability inherited from the pre-conditioning technique.

\section{\label{sec:app_vqe}VQE using Hardware-Efficient ansatz}
In this section, we show how DBQAxVQE performs with a hardware-efficient ansatz targeting the $\XXZ$ Hamiltonian. 
This ansatz is, in principle, more expressive of the Hamming-weight preserving ansatz of Fig.~\ref{fig:hwp}, in the sense that it could cover the whole Hilbert space if the chosen circuit is expressive enough. 
On one hand, this means the optimal solution to our problem lies in the search space of our algorithm.
On the other hand, however, the training is more difficult because no prior knowledge is exploited as it is when choosing the Hamming-weight preserving ansatz. 
Practically, we are also exploring regions of the Hilbert space that don't contain the target solution.
Comparing the gates composing our chosen HEA circuit with the one introduced in Fig.~\ref{fig:hwp}, we note the former presents more parameters but a lower number of CZ. The choice of the ansatz must take into account the hardware availability and calibration accuracy.

\begin{figure}[ht]
    \centering
    \includegraphics[width=0.49\linewidth]{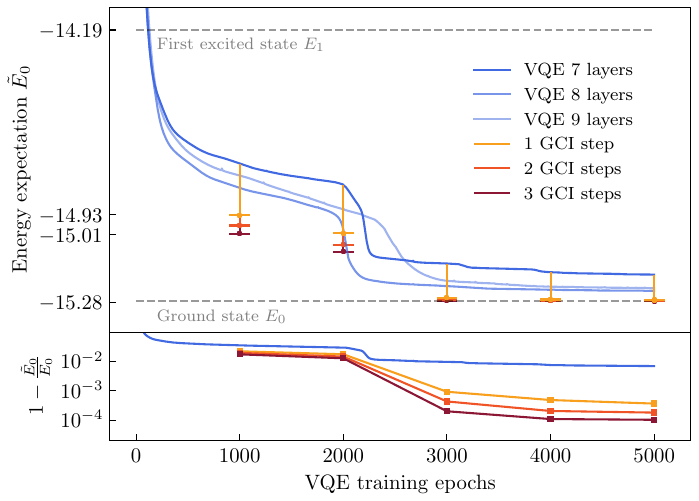}
    \includegraphics[width=0.48\linewidth]{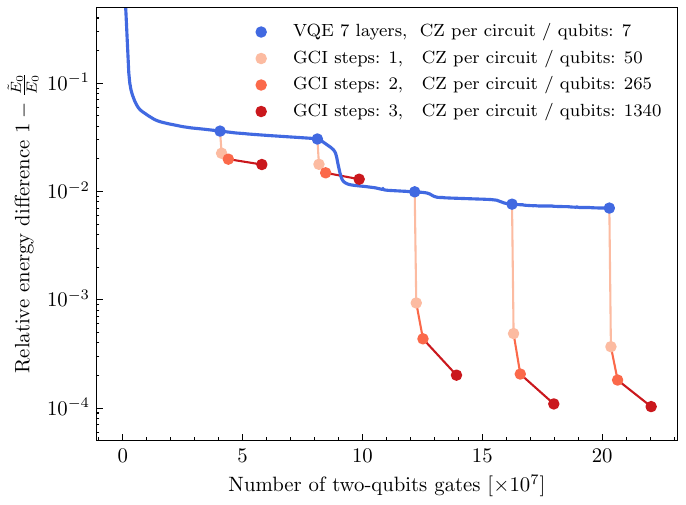}
    \caption{One example of DB-DOI~for $\XXZ$~ using a hardware efficient ansatz and obtained by fixing the simulation random 
	seed; the image is intended to provide qualitative information about the 
	impact of DB-DOI. A more robust study of the performance 
	is presented in Tab.~\ref{tab:xxz_results}.\emph{(left)} Training of VQE (blue lines) 
	for $7$, $8$, and $9$ layers (hues of blue) achieved ground state energy residue of about 1\% within $500$ training epochs.
	We initialize DBQA with VQE for selected epochs $\in [1000, \, 2000, \, 3000, \, 4000, \, 5000]$ where we apply a DBQA optimized in its parameters with CMA-ES~\cite{cma}. 
	\emph{(right)} Token cost estimates of DB-DOI~by counting the total number of two-qubit gates required to execute the complete protocol: training the VQE until a target epoch and then optimizing and applying the DBQA.}
    \label{fig:xxz_hea}
\end{figure}

\renewcommand{\arraystretch}{1.5}
\begin{table*}[ht]
\footnotesize
\begin{tabular}{c|cc|cc|cc|cc}
\hline \hline
\textbf{Layers} & \multicolumn{2}{c|}{\textbf{Warm-start}} & \multicolumn{2}{c|}{\textbf{1 DB-DOI step}} & \multicolumn{2}{c|}{\textbf{2 DB-DOI steps}} &\multicolumn{2}{c}{\textbf{Long VQE training}} \\
\hline
 & \multicolumn{2}{c|}{$1-\tilde{E}_0/E_0$} & \multicolumn{2}{c|}{$1-\tilde{E}_0/E_0$} & \multicolumn{2}{c|}{$1-\tilde{E}_0/E_0$} &\multicolumn{2}{c}{$1-\tilde{E}_0/E_0$} \\ 
\hline
 $7$ & \multicolumn{2}{c|}{$0.023\pm0.002$} & \multicolumn{2}{c|}{$0.013\pm0.0007$} & \multicolumn{2}{c|}{$0.011\pm0.0003$}  &\multicolumn{2}{c}{$0.006\pm 0.002$ } \\
 $8$ & \multicolumn{2}{c|}{$0.005\pm0.003$} & \multicolumn{2}{c|}{$0.0003\pm0.0002$} & \multicolumn{2}{c|}{$0.00013\pm0.00007$} &\multicolumn{2}{c}{$0.004\pm0.002$} \\
 $9$ & \multicolumn{2}{c|}{$0.005\pm0.003$} & \multicolumn{2}{c|}{$0.0003\pm0.0002$} & \multicolumn{2}{c|}{$0.00018\pm0.00006$} &\multicolumn{2}{c}{$0.0030\pm0.0003$} \\
\hline 
 &  Depth & Cumulative cost &  Depth & Cumulative cost &  Depth & Cumulative cost & Depth & Cumulative cost \\
\hline
$7$ & $7$ & $12.18\times 10^{7}$ & $50$ & $12.25\times 10^{7}$ & $265$ & $12.52\times 10^{7}$ &  $7$ & $20.03\times 10^{7}$  \\
$8$   & $8$ & $15.84\times 10^{7}$ & $55$ & $15.91\times 10^{7}$ & $290$ & $16.21\times 10^{7}$ & $8$ & $26.4\times 10^{7}$  \\
$9$  & $9$ & $19,98\times 10^{7}$ & $60 $ & $20.06\times 10^{7}$ & $315$ & $20.39\times 10^{7}$ &  $9$ & $33.3\times 10^{7}$ \\
\hline
\hline
\end{tabular}
\caption{\label{tab:xxz_hea_results} \textit{(above)} Energy approximation ratio for the $\XXZ$~ model of Eq.~\eqref{eq:XXZ_H} with $L=10$ qubits, and $\Delta = 0.5$. The estimates with their uncertainties were calculated using the median and the median absolute deviation of a sample of results obtained by repeating the execution fifty times with different initial conditions.
\textit{(below)} Circuit depth expressed as number of CZ gates per qubit, alongside with cumulative number 
of CZ gates used to reach $\tilde{E}_0$ (See App.~\ref{sec:training_setup}). 
Warm-start VQE approximations (3000 epochs of training) are presented alongside DB-DOI~results, executed considering compiled DB-DOI circuits. 
 Longer VQE training (5000 epochs) is reported in the last column of the table.}
\end{table*}

We proceed with the same strategy presented in the main text, but considering $N=5$ trainings per configuration. 
In Fig.~\ref{fig:xxz_hea} we show an example of DB-DOI execution with a fixed random seed. 
After an initially rapid cost function decrease, the VQE training saturates to essentially marginal improvements regardless of the number of layers considered.
On the other hand, we find that the DBQA halves the energy residue if applied to very early training epochs ($1000,\,2000$) and essentially reaches the ground state when executed later in the process.
    
The lower left plot shows that DBQA was initialized in the first training plateau is in the basin of attraction for the ground state fixed point, where convergence is exponentially fast in the number of steps.

The plot on the right shows the cumulative number of CZ gates required to execute the DB-DOI~protocol together with the circuit depth, expressed as the number of CZ gates per qubit.
We find that the training cost is dominated by VQE if considering a couple of DBQA steps. On the other hand, the DBQA training optimizes the energy more with fewer queries.
For tasks requiring high ground state preparation fidelity, both methods should be used in sequence as individually each would necessitate unnecessarily large token expenditures.
A more detailed analysis of the performances is presented in Tab.~\ref{tab:xxz_hea_results} and Tab.~\ref{tab:xxz_hea_fid}, where the approximation accuracy is respectively quantified in terms of relative differences of Eq.~\eqref{eq:reldiff} and fidelity lower bound of Eq.~\eqref{eq:fid_approx}.

\begin{table*}[ht]
\footnotesize
\begin{tabular}{c|c|c|c|c}
\hline \hline
\textbf{Layers} & \textbf{Warm-start} & \textbf{1 DB-DOI step} & \textbf{2 DB-DOI steps} & \textbf{Long VQE training}  \\
\hline
7 & $0.67\pm 0.03$ & $0.82\pm0.05$ & $0.85 \pm 0.05$ & $0.92\pm0.03$ \\
8 & $0.93\pm 0.04$ & $0.995\pm0.002$ & $0.998 \pm 0.001$ & $0.94\pm0.03$ \\
9 & $0.93\pm 0.02$ & $0.996\pm0.001$ & $0.998 \pm 0.001$ & $0.96\pm0.004$ \\
\hline
\hline
\end{tabular}
\caption{\label{tab:xxz_hea_fid} Fidelity lower bound $\mathcal{F}$ (see Eq.~\eqref{eq:fid_approx}) for the $\XXZ$~ model of Eq.~\eqref{eq:XXZ_H} with $L=10$ qubits, and $\Delta = 0.5$ extending results of Tab.~\ref{tab:xxz_hea_results}.}
\end{table*}

\section{Details about numerical simulations}
\label{sec:training_setup}
In this section, we discuss the choices we made in carrying out the simulations that led to the results shown in this manuscript. After detailing the numerical simulations, we describe the computational cost of the VQE and DB-DOI~approaches, which are reported in the tables of this work.

 All the simulations have been performed using \texttt{Qibo}~\cite{Efthymiou_2021,
Carrazza_2023, Efthymiou_2022, Efthymiou_2024, pasquale2024statevectorsimulationqibo, pedicillo2024opensourceframeworkquantumhardware, pasquale2024opensourceframeworkperformquantum}, an open-source quantum computing 
framework widely used to run quantum machine learning algorithms both in 
simulation~\cite{P_rez_Salinas_2021, Bravo_Prieto_2022, robbiati2023determining, 
Cruz_Martinez_2024, Bordoni_2023, qan_LHC, delejarza2024} and on quantum hardware~\cite{robbiati2022qadam, 
robbiati2023rtqem, app14041478}. 
The optimizations have been performed through the \texttt{Qibo} interface, which integrates robust Python based frameworks such as \texttt{keras}~\cite{chollet2015keras}, \texttt{tensorflow}~\cite{tensorflow2015-whitepaper}, \texttt{scipy}~\cite{2020SciPy-NMeth} and \texttt{pycma}~\cite{cma}.

\subsection{\label{sec:vqe_details}VQE training and computational cost}

\textbf{Training description ---} We train the VQE using a gradient-based optimization approach. We use the Adam optimizer~\cite{kingma2017adam}, which has been proven to be one of the most effective optimizers when training big machine-learning models. We set the learning rate to $0.05$ after performing a hyper-optimization on a grid of values between $0.1$ and $0.001$. We keep the default values of the remaining hyper-parameters according to the Keras' implementation.

The explored VQE architectures are the Hamming weight preserving ansatz introduced in Fig.~\ref{fig:hwp} and a hardware-efficient ansatz composed of $\operatorname{RY}$, $\operatorname{RZ}$, and $\operatorname{CZ}$ gates. The target Hamiltonians have been chosen as cases of the general Heisenberg Hamiltonian: a first target involves nearest neighbors interactions and a penalty $\Delta=0.5$ to the $\hat{Z}$s, while in a second moment we consider a more complex case lighting up the next-nearest neighbors interactions.
We refer to these two cases as $\XXZ$~ and $J_1$-$J_2$ respectively.

Once the VQE ansatz and the target Hamiltonian are specified, we explore its expressibility by varying the number of layers in the VQE ansatz. 
We trained both ansatze with layers ranging from $3$ to $9$, and we repeated each training with five different initial configurations. Namely, we randomly sampled the initial values of the angles parametrizing the variational model from the uniform distribution $\mathcal{U}(-\pi, \pi)$. 

After selecting the circuit sizes for which the ground state approximations were most accurate, we conducted a series of more exhaustive simulations to quantify the training error. In particular, we handle the $\XXZ$~ target training Hamming-weight preserving ansatze composed of 3, 4, and 5 layers for $2000$ epochs, and hardware-efficient ansatze composed of 7, 8, and 9 layers for $5000$ epochs. We instead tackle the $J_1$-$J_2$ target training Hamming-weight preserving ansatze composed of 3, 4, 5, and 6 layers for $2000$ epochs.

For each fixed target, ansatz, and number of layers, we repeat the training $50$ times, with a different set of initial parameters sampled from the uniform distribution $\mathcal{U}(-\pi, \pi)$.
Each of these training instances corresponds to a ground state energy approximation $\tilde{E}_0$, whose quality is quantified by computing the relative difference with the target (known or numerically computed) ground state energy $E_0$
\begin{equation}
\text{RD} = 1 - \frac{\tilde{E}_0}{E_0}.
\label{eq:reldiff}
\end{equation}
Alternatively, we can use the ground state energy approximations $\tilde{E}_0$ to compute the following fidelity lower bound~\cite{cramer2010efficient}
\begin{equation}
    \mathcal{F} = 1 - \frac{\tilde{E}_0 - E_0}{E_1 - E_0},
    \label{eq:fid_approx}
\end{equation}
where $E_0$ and $E_1$ are the true ground state and first excited state, respectively.

Once all the fifty RD or $\mathcal{F}$ values are collected, a final estimate is computed through the median value of the list, $\text{median}(\bm{x})$,
where we call $\bm{x}$ the list of RD/$\mathcal{F}$ values for simplicity.
The estimation of the uncertainty is instead computed using the median absolute deviation:
\begin{equation}
\text{MAD} = 1.4826 \cdot \text{median}\bigl( | x_i - \text{median}(\bm{x}) | \bigr).
\label{eq:mad}
\end{equation}
We choose the median and the median absolute deviation as estimators of the variable and its uncertainty to be more robust to outliers, which could be expected when applying the DB-DOI~ algorithm. Since we make use of a global cost function in this work, it can happen that, if the initial approximation provided by the VQE is not close enough to the ground state, the final effect of the DBQA doesn't correspond to a further reduction of the energy. Moreover, since the optimization cost of the DBQA is particularly intense in the case of the non-compiled DBI, in some rare cases it can happen the optimizers are not able to find an optimal configuration of the parameters when the optimization process is limited in time.

\vspace{1.5mm}
\textbf{Computational cost ---} To evaluate the computational cost of the VQE training, we need to take into account the total number of two-qubit gates composing the circuit architecture and the number of times an expectation value has to be computed to evaluate predictions and gradients during the training. When using a gradient-based approach on a quantum device, the gradients must be calculated using parameter shift rules~\cite{Schuld_2019, Mitarai_2018}. Since we parametrize the quantum circuit through rotational gates, it is well known the partial derivative of our cost function w.r.t. to a circuit's parameter $\theta$ can be calculated using two expectation values. Considering the choice of Adam optimization, the entire gradient of the cost function has to be computed at each optimization iteration. Finally, the total amount of two-qubit gates (we take CZ as reference) can be computed as:
\begin{equation}
N_{\rm CZ}^{\rm VQE} (e) = k \cdot p \cdot e \cdot n_{\rm CZ}^{\rm VQE},
\label{eq:cz_count}
\end{equation}
where we indicate with $e$ the number of Adam iterations, $p$ the number of parameters of the circuit, and $n_{\rm CZ}^{\rm VQE}$ the number of CZ gates that compose the circuit according to the chosen ansatz. 
The constant $k$ refers to the number of expectation values required to execute the parameter shift rule. In the case of the hardware-efficient ansatz $k=2$ and in the case of the Hamming weight preserving ansatz $k=4$, since each RBS gate is decomposed into two rotations depending on the same angle $\theta$~\cite{Cherrat_2024}. 
According to the same decomposition, the number of CZ gates composing the Hamming-weight preserving ansatz circuit can be calculated as twice the number of RBS gates. 

\subsection{DB-DOI~procedure and computational cost}

\textbf{DB-DOI~procedure ---} To fully exploit the algorithm, it is necessary to prepare an approximation of the ground state. In this work, we make use of a VQE, but any ground state preparation algorithm can be used. This first approximation is then used to precondition the target Hamiltonian, which is then rotated according to Eq.~\eqref{vqaxdbqa}. The following key step consists in compiling the DBI circuit and this can be done following an equivalent procedure to the one presented in Sec.~\ref{sec:xxz_compiling}. In the following discussion of the computational cost of the algorithm, we will take into account the same VQE trainings exposed in the previous section, which are used as preconditioning ground state approximations in the DB-DOI~process. We evaluate Eq.~\eqref{eq:reldiff} after applying the DB-DOI~to a given ground state approximation provided by VQE stopped at target epoch $e$. Following the same procedure of Sec.~\ref{sec:vqe_details} we collect all the RD values and provide the final estimations as the median and median absolute deviation of the obtained results.

\vspace{1.5mm}
\textbf{Computational cost ---} Both the initial state approximation and the DB-DOI compilation into a circuit present a computational cost and have to be cumulatively taken into account to evaluate the whole computational expense of the algorithm. The total number of two qubits gates required by the DB-DOI~process involves then a first contribution due to the VQE cost described in Eq.~\eqref{eq:cz_count}. We have to consider then the cost of executing the new circuit, in which the VQE unitary is repeated exponentially as DBI iterations increase, and there is an additional cost due to the Hamiltonian decomposition as explained in Sec.~\ref{sec:xxz_compiling}. Denoting $n_{\rm CZ}^{\rm DBQA}$ the number of CZ gates composing the circuit already involving both the DB-DOI compilation and the recursive VQE call, we finally must consider the cost of optimizing the DB-DOI's parameters. In fact, the DB-DOI circuit can be parametrized both in the step duration $s$ and in the diagonal operators $\D_k$. In this work, we parametrize $\D_k$ as a classical Ising model with nearest-neighbor (NN) interactions:
\begin{align}
\D_k(B^{(k)},J^{(k)})=\sum_{i=0}^N(\alpha_i^{(k)}\Z_i + \beta_i^{(k)}\Z_{i+1}\Z_i),
\end{align}
where $N$ is the number of sites in the chain, the parametrization is implemented through the coefficients $\{\alpha_i\}_{i=0}^N$ and $\{\beta_i\}_{i=0}^N$, and the superscript $k$ highlights that the optimization procedure is repeated for each DBQA step.
We use Scipy's Powell and CMA-ES optimizers to find an optimal DBI configuration, and this optimization has to be considered as an additional cost in the overall count.
To evaluate this final contribution to the total number of two-qubit gates we multiply $n_{\rm CZ}^{\rm DBQA}$ with the total number of cost function evaluations $n_{\rm fval}$ executed by the optimizers. 
The total number of two-qubit gates required to finalize the DB-DOI~ground state approximation can be finally obtained as
\begin{equation}
N_{\rm CZ}^{\rm DBQA} = N_{\rm CZ}^{VQE}(e) + n_{\rm fval} \cdot n_{\rm CZ}^{\rm DBQA}.
\label{eq:dbqa_cz_cost}
\end{equation}

\section{Compiling of \textup{XXZ} evolution}
\label{sec:xxz_compiling}

We next provide a discussion on how to explicitly express the diagonalization DBQA as above into an explicit circuit.
For the $\XXZ$~ model, we write the Hamiltonian as 

\begin{align}
	\h_0 = \sum_{a=1}^{L} \h^{(a)}
	\label{h_compiling}
\end{align}
where each summand addresses only two qubits
\begin{align}
    \h^{(a)} = \X_a \X_{a+1}+\Y_a \Y_{a+1}+\Z_a \Z_{a+1} 
\end{align}
for $a<L$ and 
\begin{align}
    \h^{(L)} = \X_L \X_{1}+\Y_L \Y_{1}+\Z_L \Z_{1}
 \end{align}
 because of periodic boundary conditions.
We next use $M$ steps of the linear Trotter-Suzuki decomposition
\begin{align}
	e^{-it \h_0} = \left( \prod_{a=1}^L e^{-i\frac{t}{M}\h^{(a)}}\right)^M+O(M^{-1})
\end{align}
which means that, if we decompose $e^{-i t/M \h^{(a)}}$ into CNOT and single qubit rotations, we get a circuit approximation with accuracy $O(t^2/M)$.
This is done by noticing that $e^{-i t/M \h^{(a)}}$ is a unitary acting on two qubits, \emph{i.e.} $a$ and $a+1$, and any two-qubit unitary can be decomposed into a circuit made of single qubit rotations and 3 CNOT gates~\cite{VatanWilliams}.

We next discuss generalizations.
The Hamiltonian in Eq.~\eqref{h_compiling} can be arbitrary and only the last step has to be modified: Then the terms $\h^{(a)}$ can be acting on more than 2 qubits and the evolutions they generate $e^{-i t/M \h^{(a)}}$ can be compiled into CNOTs and single qubit rotations using the quantum Shannon decomposition~\cite{shende2005synthesis} which implements a unitary on $K$ qubits using $O(4^K)$ CNOTs which is, in general, an optimal scaling.

Additionally, we can use a higher-order Trotter-Suzuki decomposition.
We assume again that we have a two-qubit Hamiltonian on a line of an even number of qubits $L$ so that $\h_0^\text{(o)} = \sum_{a=1}^{L/2} \h^{(2a-1)}$ and $\h_0^\text{(e)}$ contains only commuting terms.
Then we write
\begin{align}
	e^{-it \h_0} = \left( e^{-it/(2M)h_0^\text{(o)}}e^{-it/M\h_0^\text{(e)}}e^{-it/(2M)\h_0^\text{(o)}}  \right)^M+O(M^{-2})
\end{align}
which gives an improved error scaling.
From commutativity, we have the exact equality
\begin{align}
	 e^{-it/(2M)\h_0^\text{(o)}}= \prod_{a=1}^{L/2} e^{-i t/M \h^{(2a+1)}}
\end{align}
Each of the factors above can again be decomposed into CNOTs and single-qubit rotations by standard methods~\cite{VatanWilliams,shende2005synthesis}.

\subsection{Other 2-local models}

The addition of the magnetic field Hamiltonian can be compiled in the above formalism
by setting
\begin{align}
    \h^{(a)} = \X_a \X_{a+1}+\Y_a \Y_{a+1}+\Z_a \Z_{a+1} + B_a Z_a
\end{align}
for $a<L$ and for periodic boundary conditions, we set 
\begin{align}
    \h^{(L)} = \X_{L} \X_{1}+\Y_{L} \Y_{1}+\Z_{L} \Z_{1}+B_L\Z_L
 \end{align}
 or  for open boundary conditions $\h^{(L)}=0$ so that
\begin{align}
	\h_0 = \h_{\XXZ} + \h(B) = \sum_{a=1}^{L} \h^{(a)}\ .
\end{align}
For the transverse-longitudinal field Ising model, we set
\begin{align}
    \h^{(a)} = \X_a \X_{a+1}+ B_a \Z_a+C_a \X_a
\end{align}
and boundary terms similar to those above.
This would give as above a compiling solution with 3 CNOT gates.

\subsection{Special-purpose compiling for the transverse-field Ising model}
We next consider the special transverse-field Ising model
\begin{align}
    \h^{(a)} = \X_a \X_{a+1}+ B_a \Z_a
\end{align}
obtained from the above by setting $C_a=0$.
We use that $\X_a \X_b = \text{CNOT}(a,b) \X_a \text{CNOT}(a,b)$ and $\Z_a  = \text{CNOT}(a,b) \Z_a \text{CNOT}(a,b)$ to write
\begin{align}
	\text{CNOT}(a,b)\h^{(a)} \text{CNOT}(a,b)= \X_a + B_a \Z_a
\end{align}
which means that we can get for any $t$
\begin{align}
	e^{-it\h^{(a)} } = \text{CNOT}(a,b) e^{-it(\X_a + B_a \Z_a)}	\text{CNOT}(a,b)\ .
\end{align}
In other words, the transverse-field Ising model evolutions can be compiled using 2, not 3, CNOT gates per interaction term.
\subsection{Compiling for the classical Ising model}
In the numerical calculations, we use parametrizations for the diagonal evolutions that have a low quantum compiling.
More specifically, we define the classical Ising model
\begin{align}
	\h(B,J) = \sum_{a=1}^L(B_a Z_a +J_{a,a+1} \Z_a \Z_{a+1})
\end{align}
where as above $Z_{L+1}=Z_1$.
For $\h(B,J)$, we can compile the diagonal evolution using 2 CNOT gates, which is more efficient than using the general method of compiling two-qubit unitaries from Ref.~\cite{VatanWilliams}.
We use that $Z_a Z_b = \text{CNOT}(a,b) Z_b \text{CNOT}(a,b)$ and by unitarity of the CNOT gate
\begin{align}
	e^{-it Z_a Z_b} = \text{CNOT}(a,b) e^{-it Z_b} \text{CNOT} (a,b)\ .
\end{align}
The model is commuting so $e^{-it\h(B, J=0)}$ consists of independent single qubit rotations.

\section{DB-DOI~results considering the $J_{1}$-$J_{2}$ model}
\label{sec:j1j2}

In the main text, we presented results for the $\operatorname{SU}(2)$ symmetric $\XXZ$~ model where we set $\Delta=0.5$ as a penalty to the $\hat{Z}$ interactions.
Here, we extend the analysis to a more general model:
\begin{align}
  \h_{J_1\text{-}J_2} = 
  J_1\h_{\XXZ} + J_2 \sum_{i=1}^{L} (\X_i \X_{i+2} + \Y_i\Y_{i+2} + \Z_i \Z_{i+2}),
  \label{eq:j1j2_H}
\end{align}
where we light up next-nearest neighbors interactions. In particular, we consider $J_1=1$ and $J_2=0.2$, which corresponds to a regime such that $J_2/J_1=0.2$, and has been chosen knowing the target system presents a Berezinskii–Kosterlitz–Thouless transition at $J_2/J_1=0.24116$~\cite{Jos201340YO, PhysRevB.54.9862}. 
Also, in this case, we use the Hamming-Weight preserving ansatz presented in Fig.~\ref{fig:hw_xxz} because the same symmetries are respected. 
We follow the same procedure as the one presented in the main text for $\XXZ$~, but in this case, we execute the DBQA in single commutator mode only. 
We postpone to future works the compilation of the model into a quantum circuit, likewise we did in App.~\ref{sec:xxz_compiling}. 
The obtained results are presented in Tab.~\ref{tab:j1j2_results}.

\renewcommand{\arraystretch}{1.5}
\begin{table*}[ht]
\footnotesize
\begin{tabular}{c|c|c|c|c|c}
\hline \hline
\textbf{Layers} & \textbf{Warm-start} & \textbf{1 DBI step} & \textbf{2 DBI steps} & \textbf{3 DBI steps} & \textbf{Long VQE training}\\
\hline
$3$  & $0.033\pm0.005$  & $0.020\pm0.005$  & $0.016\pm0.005$   & $0.014\pm0.004$ & $0.026\pm0.006$  \\
$4$   & $0.017\pm0.007$  & $0.006\pm0.005$  & $0.003\pm0.003$  & $0.002\pm0.002$  & $0.010\pm0.004$ \\
$5$    & $0.011\pm 0.007$  & $0.002\pm 0.002$ & $0.0008\pm 0.0008$  & $0.0004\pm 0.0004$ & $0.005\pm 0.003$ \\
$6$   & $0.009\pm 0.006$  & $0.002\pm 0.001$ & $0.0007\pm 0.0007$  & $0.0004\pm 0.0004$ & $0.005\pm 0.003$\\
\hline \hline
\end{tabular}

\caption{\label{tab:j1j2_results}Relative difference between approximated energy $\tilde{E}_0$ and the target  ground state value $E_0$ for the $J_1$-$J_2$~ model of Eq.~\eqref{eq:j1j2_H} with $L=10$ qubits, $J_1=1$ and $J_2=0.2$ together with the cumulative number 
of CZ gates, $N_{\rm CZ}$, used to reach $\tilde{E}_0$. The estimates with their uncertainties were calculated using the median and the median absolute deviation of a sample of results obtained by repeating the execution fifty times with different initial conditions.
Warm-start VQE approximations (500 epochs of training) are presented alongside DB-DOI~results, executed considering BDI unitaries. 
Longer VQE trainings (2000 epochs) are reported in the last column of the table. }
\end{table*}

\section{Details on relation to other methods}
\label{app:comparison}

There exist many other VQE ideas, and we refer to reviews for specific discussion of their detailed performance~\cite{KishorRevModPhys.94.015004,cerezo2021variationalReview,tilly2022variationalReview}.
Overall, it is clear that purely variational methods can be limited by training obstructions (\emph{e.g.} swamps of local minima or barren plateaus). So, the accuracy in energy estimation is expected to saturate before reaching the global minimum.
In all variational cases, one can interface the unitary $\vqes$ with DBQA via the rotation of the input generator as done in the main text.

We note that DB-DOI~ is conceptually distinct from Ref.~\cite{riemannianflowPhysRevA.107.062421}, which also uses methods involving exponentials of commutators, framed there in the language of Riemannian flows. The difference is that Ref.~\cite{riemannianflowPhysRevA.107.062421} focuses on computing the gradients of generic fixed parameterized quantum circuits while we work with the full manifold rather than a submanifold. This eases implementations of state updates in gradient directions and unlike in Ref.~\cite{riemannianflowPhysRevA.107.062421} measurements are not required in DB-DOI.
We refer to Ref.~\cite{schulte2010gradient} for an early review Riemannian gradients and Riemannian gradients as an underlying principle for performing quantum control and quantum computing tasks.
Ref.~\cite{helmke_moore_optimization,moore1994numerical} studied mathematically iterations involving exponentials of commutators under the name Lie-bracket recursions (while we refer to them as double-bracket iterations for consistency with double-bracket flows referring to the continuous flows) and proved their convergence. 

 Gradient-based optimization is often used for training VQE, and it is natural to consider these methods for training DBQA as well. 
An example of such $1$-dimensional optimization is to use a greedy strategy to select DBI step durations $s_k$ as cost-function minimizers for the respective DBR. 
This strategy can be more general, \emph{e.g.} gradient descent in the space of magnetic fields $B_i$ or Ising couplings $J_{i,j}$ to find the best $\D_k$ operators in each step, starting from an initial guess.
We have also tried non-gradient-based optimizers, see App.~\ref{sec:training_setup}.

An interesting avenue of research is to replace VQE unitaries $\vqe$ by Hartree-Fock warm-starts, see Ref.~\cite{arute2020hartree} for a discussion of quantum compiling using Givens rotations and experimental data.
In this case, again, the question would be whether Hartree-Fock interfaced with DBQA could achieve better ground state approximation.
Further generalization to other ansatze in quantum chemistry can be interfaced with DBQA as long as explicit compiling is available.

Apart from direct minimization of energy, there exist methods that aim to implement certain transformations, and these transformations have energy-lowering properties.
Quantum imaginary-time evolution (QITE) is an iteration where in each step one seeks an approximation to the imaginary-time update $\ket{\psi_{k+1}} = e^{-\tau \h_0}\ket{\psi_{k}}/ \|e^{-\tau \h_0}\ket{\psi_{k}}\|$ through a unitary $\u_k$ such that $\u_k \ket{\psi_{k}} \approx \ket{\psi_{k+1}}$.
Here, the optimal parametrization of $\u_k$ is not known so variational methods have been used~\cite{mcArdle2019, avqite, Motta2020}.
QITE circumvents trainability issues by trying to implement steps of an iteration that is bound to decrease the energy, but it is not clear how to find the correct unitary for longer durations where the cost function resolution is limiting.
Instead, DBQA is coherent in that it can be applied directly and only relies on measurements to optimize its performance but does not rely on classical supervision for its parametrization.
For this reason, it is likely that the QITExDBQA interfacing should be preferred over the DBQAxQITE.
However, if it is possible to easily parameterize QITE at late stages of energy optimization, then QITExDBQAxQITE could be useful too.

So far, we discussed methods that have been suggested without the use of auxiliary qubits.
If one allows for that, then methods based on quantum phase estimation (QPE) become key.
Here again, interfacing can be possible.
Most likely DBQAxQPE should be preferred over QPExDBQA because QPE can be expected to be more costly than DBQA, so it is natural to chain methods to increase resource cost, \emph{e.g.} VQExDBQAxQPE.
QPE methods can be systematically formulated in the formalism of qubitization.
Note that QITE can also be implemented in this way. In all cases, the upper bound on the ground state energy found via DB-DOI~ (or DBQAxQPE) could also be complemented with a lower bound of the ground state energy via Dual-VQE~\cite{westerheim2023dual}.

Finally, let us comment broadly on DBQA concerning the so-called \emph{quantum approximate optimization algorithm} (QAOA)~\cite{kokail2019self,KishorRevModPhys.94.015004}.
Let us begin with Eq.~\eqref{eq:U1DBDOi} and notice the resemblance to a $1$-layer QAOA, in that in $\hat U_1$ we have a layer of single-qubit gates (specifically, diagonal $Z$-rotations) followed by the evolution under the input Hamiltonian and then again single-qubit rotations.
For the first step, this is similar to QAOA which interlaces single-qubit rotation layers with evolutions under the input Hamiltonian.
The biggest difference is that in DBQA, the two single-qubit layers are related in that they are inverses of each other.
Thus, the guidance of the DBI equation through the reduced group commutator approximation suggests a more constrained QAOA ansatz.
This constraint facilitates the monotonicity relation~\cite{Gluza_2024}, which allows us to analytically understand why diagonalization ensues.
Similar relations for unconstrained QAOA are not known to us, but Ref.~\cite{Gluza_2024} conjectured that this might hint at understanding the functioning of optimized QAOAs.
Aside from these coincidental similarities, the methods are different in several key aspects.
In Ref.~\cite{kokail2019self}, the QAOA ansatz involving evolutions with the native Hamiltonian for trapped ions was used to prepare an approximation to the ground state of a different model.
In contrast, DBQA is not restricted to target only the ground state but can be applied to target any other eigenstate of the given model.
The distinction from QAOA becomes more evident when applying DBQA beyond the first step.
For example, in the second step,\eqref{equnfoldedG1} another difference appears in that $\v_2$ involves not only forward evolutions under the input Hamiltonian but also respective backward evolutions.
Indeed, while it may be challenging to implement DBQA beyond the first step on analog quantum simulators, it is this nested forward and backward evolution structure that ensures the convergence of DBQA.
We are not aware of convergence guarantees of the DBQA type for QAOA.

\subsection{Overview of currently available circuit depth}
Following the overview above, most methods for preparing ground states on existing quantum hardware suffer from one of two severe problems.
Variational methods are limited by the resolution of the cost function, which is at the root of their operation.
This limits the circuit depth that is meaningful in practice to about a dozen entangling layers.
QPE-based methods are the opposite in that to become meaningful, they need a large circuit depth.

As a specific example, we consider Ref.~\cite{quantinuumQPE}, which compared the performance of QPE for physical and logical qubits.
The algorithmic performance in Ref.~\cite{quantinuumQPE} was obtained by using $N \approx 920$ CZ gates for $L=8$ qubits, and the two-qubit gate fidelity was $p_e \approx 2\times 10^{-3}$.

We can use these experimental results to get ballpark figures for currently available depths.
Upcoming devices are set to reach $p_{e}^{\prime} \approx 5\times 10^{-4}$~\cite{kentaro_private_com}, so assuming each CZ gate fails independently of the others, we can estimate the number $N^{\prime}$ of gates that can be used meaningfully by solving heuristically
\begin{align}
     (1-p_{e})^{N} = (1-p_{e}^{\prime})^{N^{\prime}}
\end{align}
This equation gives $16\%$ success probability, which in the three-nines fidelity regime $p_{e}^{\prime}$ would appear for $N^{\prime} \approx 3690$ CZ gates.
From these estimates, we see that there exists quantum hardware which operates meaningfully for circuit depths of about $100$ CZ gates per qubit, and this is set to even larger values in the very near future.
This means that $1$ and $2$ DB-DOI steps are already feasible for implementation on existing noisy intermediate-scale quantum devices.

\end{document}